\documentclass[%
 aip,
 amsmath,amssymb,
]{revtex4-1}

\usepackage{graphicx}
\usepackage{dcolumn}
\usepackage{bm}

\usepackage[utf8]{inputenc}
\usepackage[T1]{fontenc}
\usepackage{mathptmx}
\usepackage{etoolbox}

\makeatletter
\def\@email#1#2{%
 \endgroup
 \patchcmd{\titleblock@produce}
  {\frontmatter@RRAPformat}
  {\frontmatter@RRAPformat{\produce@RRAP{*#1\href{mailto:#2}{#2}}}\frontmatter@RRAPformat}
  {}{}
}%
\makeatother
\begin{document}

\preprint{Arxiv version}

\title[]{Aerodynamic bag breakup of a polymeric droplet}
\author{Navin Kumar Chandra}
\author{Shubham Sharma}%
 \affiliation{Department of Mechanical Engineering, Indian Institute of Science, Bangalore- KA560012, India}

\author{Saptarshi Basu*}
 
 \email{sbasu@iisc.ac.in}
\affiliation{Department of Mechanical Engineering, Indian Institute of Science, Bangalore- KA560012, India}
\affiliation{Interdisciplinary Centre for Energy Research, Indian Institute of Science, Bangalore- KA560012, India}

\author{Aloke Kumar*}
 
 \email{alokekumar@iisc.ac.in}
\affiliation{Department of Mechanical Engineering, Indian Institute of Science, Bangalore- KA560012, India}

\begin{abstract}
The aerodynamic breakup of a polymeric droplet in the bag breakup regime is investigated experimentally and compared with the result of the Newtonian droplet. To understand the effect of liquid elasticity, the Weber number is kept fixed ($\approx$ 12.5) while the elasticity number is varied in the range of $\sim 10^{-4}-10^{-2}$. Experiments are performed by allowing a liquid droplet to fall in a horizontal, continuously flowing air stream. It is observed that the initial deformation dynamics of a polymeric droplet is similar to the Newtonian droplet. However, in the later stages, the actual fragmentation of liquid mass is resisted by the presence of polymers. Depending upon the liquid elasticity, fragmentation can be completely inhibited in the timescale of experimental observation. We provide a framework to study this problem, identify the stages where the role of liquid elasticity can be neglected and where it must be considered, and finally, establish a criterion that governs the occurrence or the absence of fragmentation in a specified time period.
\end{abstract}

\maketitle

\section{Introduction}
\label{sec:introduction}

Fragmentation of a liquid bulk into smaller units is a fascinating and one of the most important fundamental topics in fluid mechanics. It has been studied for a long time, for instance, the seminal work of Plateau (1873) \cite{eggers2008physics} on the disintegration of a descending liquid jet, which laid the groundwork for the well-known Rayleigh-Plateau instability. Despite its long history of inquiry, the topic of liquid fragmentation still remains an active area of research. In this context, \citet{villermaux2007fragmentation} correctly pointed out that continued exploration in this field should not be interpreted as a lack of progress. Instead, it reflects the importance of the topic and its relevance in novel and emerging applications such as inkjet printing for additive manufacturing \citep{lohse2022fundamental}, powder production for metal 3D-printing \citep{sharma2023shock}, atomization of gelled fuel in rocket engines \citep{padwal2021gel}, and understanding the mechanism of disease transmission via parcels of micron-sized liquid droplets \citep{sharma2021secondary, scharfman2016visualization}. Aerobreakup is one particular method of achieving liquid fragmentation wherein the liquid mass is subjected to a high-speed stream of gas (generally air), causing the liquid mass to disintegrate into smaller fragments. It is broadly divided into two sub-topics, the primary breakup, which involves the formation of liquid droplets from a bulk liquid mass, and the secondary breakup in which a liquid droplet further breaks up into smaller fragments. The present work is focused on the secondary breakup of a polymeric droplet that exhibits viscoelastic properties. Compared to non-Newtonian droplets, the secondary breakup of a Newtonian droplet is researched more extensively and well-reviewed in the literature \citep{pilch1987use,faeth1995structure,gelfand1996droplet,guildenbecher2009secondary,theofanous2011aerobreakup,sharma2022advances}. In the aerobreakup process, the aerodynamic force plays the disruptive role while the liquid viscosity, inertia, surface tension, and elasticity (if present) provide resistance against the breakup. Through extensive experiments along with support from theoretical and numerical analysis, it is now well-established that the Weber number ($We$) and the Ohnesorge number ($Oh$) are the two most important non-dimensional parameters governing the aerobreakup of a Newtonian droplet. These numbers are defined as follows-
\begin{equation}
We=\frac{\rho_g U_g^2 D_0}{\gamma}
\label{eqn1}
\end{equation}
\begin{equation}
Oh=\frac{\mu_l}{\sqrt{\rho_l \gamma D_0}}
\label{eqn2}
\end{equation}
Here, $\rho_g$ and $\rho_l$ are the gas and the liquid phase density, $\gamma$ is the surface tension, $\mu_l$ is the dynamic viscosity of the liquid phase, $U_g$ is the free stream velocity of the gas phase, and $D_0$ represents the initial diameter of the droplet. $We$ represents the ratio of aerodynamic force to the surface tension force, and it is equally important for Newtonian as well as non-Newtonian droplet aerobreakup. The effect of liquid viscosity is reflected in the $Oh$, which is simple to calculate for Newtonian liquids; however, it is not straightforward for non-Newtonian droplets due to their possibility of deformation rate-dependent viscosity. Depending upon the $We$, secondary breakup happens through different modes. In the increasing order of their $We$ range, these modes are- vibrational or no breakup, bag, bag-stamen, multi-mode, shear-stripping, and catastrophic breakup modes \citep{hsiang1992near,hsiang1995drop}. For most liquids of practical purpose, the liquid viscosity does not play any significant role, and the breakup mode is decided by the $We$ alone. Particularly, if criteria of $Oh<0.1$ (from hereon referred to as low-viscosity droplet) is satisfied, then the liquid viscosity does not play any role in deciding the breakup mode \citep{guildenbecher2009secondary}. \citet{theofanous2013physics} showed that the same criteria could be extended to the non-Newtonian droplets if the $Oh$ is calculated based on an effective viscosity of the liquid phase accounting for the relevant deformation rate involved in the process. A similar observation is corroborated by our previous work \citep{chandra2023shock}, where we showed that up to very large extents, the presence of liquid elasticity does not play any significant role in deciding the mode of aerobreakup.

Among all the breakup modes mentioned above, the bag breakup is the most important mode and has received the highest attention \citep{chou1998temporal,flock2012experimental,jalaal2012fragmentation,kulkarni2014bag,opfer2014droplet,zhao2018transition,soni2020deformation,jackiw2021aerodynamic,jackiw2022prediction,qian2021experimental,jain2019secondary,xu2023transitions,zhao2010morphological}. The bag breakup mode for low-viscosity droplets is obtained for $We$ ranging from $\sim$11 to $\sim$16 \citep{kulkarni2014bag} with slight variation in the exact $We$ value reported in different literature. This mode is most important because it separates the regime of no breakup from the onset of the breakup, i.e., there is no further breakup below the range of $We$ for bag breakup mode. The importance of the bag breakup mode is evident from the fact that the physics of a single droplet bag breakup has the power to explain the peaks obtained in the fragment size distribution of many industrial sprays and the droplet size distribution in the falling raindrops \citep{villermaux2009single,kulkarni2014bag}.

Polymeric droplets exhibiting viscoelastic properties draw special attention because of two reasons. First, many of the liquids of practical importance are inherently viscoelastic, and their breakup can be drastically different from the Newtonian liquids, for instance, the breakup of salivary droplets during sneezing and spray drying of fruit pulps \citep{scharfman2016visualization,cervantes2014study}. Second, small amounts of polymers can be added as a rheological modifier to control the aerobreakup process. One of the best examples in this context is the spray of agricultural chemicals, where polymers are added to prevent the formation of very fine driftable fragments during the spray process \citep{mun1999atomisation}. While different aspects of a Newtonian droplet secondary breakup have been studied thoroughly in the existing literature, the studies related to the aerobreakup of viscoelastic droplets are sparse. The early research in the aerobreakup of polymeric droplets comes from the work of \citet{wilcox1961retardation} and \citet{matta1982viscoelastic,matta1983aerodynamic}. They tested various concentrations of different polymers in a very high-speed gas flow, typically 200 to 300 m/s, motivated by the use of polymers as a rheological modifier for the aircraft-assisted delivery of chemicals. These early studies concluded that the addition of polymers retards the aerobreakup, which results in higher breakup time and larger fragment size compared to their Newtonian counterpart. \citet{matta1983aerodynamic} also put forward an important conclusion that the aerobreakup phenomenon is governed by the elongational properties of the liquid rather than the shear properties. \citet{arcoumanis1994breakup} performed similar experiments and noted that the aerobreakup of a polymeric droplet happens through the intermediate formation of long ligaments, which is different from the breakup mechanism of a Newtonian droplet. In the endeavor to decipher the breakup mechanism, \citet{joseph1999breakup,joseph2002rayleigh} performed shock-induced aerobreakup experiments with viscous and viscoelastic droplets at very high Weber numbers ($\sim10^4-10^5$). The authors reported that a widespread catastrophic breakup of droplets at such high Weber numbers is assisted by the appearance of surface corrugations at an early stage on the flattened windward face of the droplet. They also proposed that the front surface corrugations are due to the Rayleigh-Taylor (RT) instability, based on the good match of wavelengths obtained from theory and the experiments. Theofanous and co-authors \citep{theofanous2008physics,theofanous2011aerobreakup,theofanous2013physics} explored the possibility of studying the Newtonian and the viscoelastic droplet aerobreakup under a single roof and provided a new way of classifying the breakup modes based on the underlying hydrodynamic instability. In this new approach, the first breakup regime observed at low $We$ is termed as Rayleigh-Taylor piercing (RTP) which is governed by RT instability and encompasses the bag, bag-stamen, and multi-mode breakup regime of traditional classification. The second breakup regime observed at high $We$ is termed shear-induced entrainment (SIE), governed by the Kelvin-Helmholtz (KH) instability, corresponding to the shear-stripping breakup mode of traditional classification. \citet{theofanous2013physics} reported that in the RTP and SIE regime, a viscoelastic droplet undergoes deformation and morphology changes the same as a Newtonian droplet, but unlike Newtonian liquids, actual fragmentation (authors used the word particulation) is not observed for viscoelastic liquids. The fragmentation of viscoelastic liquid is reported to happen at gas speeds well past the onset of the SIE regime. Therefore authors defined a new regime of shear-induced entrainment with rupture (SIER) which is applicable only to polymeric (viscoelastic) droplets. Our recent work \citep{chandra2023shock} demonstrated that liquid elasticity does not play a significant role during the early stages of droplet breakup, where droplet deformation and hydrodynamic instabilities (KH and RT) are observed. But a dominant role of liquid elasticity is observed during the final stages in terms of the morphology of the liquid mass, where large deformations at high strain rates are involved.

It is imperative to note that the ultimate aim of studying secondary breakup is to obtain the final fragment size distribution \citep{sharma2023depth,wang2018universal}. However, most of the studies, including Newtonian and non-Newtonian, focus only on identifying different breakup modes and the early deformation dynamics of the droplet, which happens before actual fragmentation during the secondary breakup process. \citet{jackiw2021aerodynamic,jackiw2022prediction} recently provided a comprehensive study on the aerobreakup (mainly focused on the bag and bag-stamen mode) of a Newtonian droplet covering all aspects starting from deformation to mode prediction and, finally, the fragment size distribution. The authors mentioned that the study of bag breakup is challenging due to the multitude of lengthscales and timescales involved in the process. The presence of liquid elasticity adds another layer of complexity, and there is no such comprehensive literature on the aerobreakup of a viscoelastic droplet. In fact, one can talk about fragment size distribution only if the fragmentation is guaranteed. While the process of fragmentation after entering a breakup regime is trivial for a Newtonian droplet, it is not the case with viscoelastic droplets \citep{arcoumanis1994breakup,theofanous2013physics}. Therefore, in the present work, we delve into finding a criterion governing the actual fragmentation to happen in the bag breakup regime of a polymeric droplet. Experiments are performed by recording high-speed images of a liquid droplet allowed to fall in a continuous airflow stream. Polymeric droplets are obtained from different concentrations of two different polyethylene oxides dissolved in a water-glycerol-based solvent. The liquid elasticity is expressed in terms of the elasticity number ($El$), which is an important non-dimensional parameter in the study of a viscoelastic liquid breakup \citep{sharma2022advances,chandra2023shock}, and defined as follows-
\begin{equation}
El=\frac{\lambda \mu_0}{\rho_l D_0^2}
\label{eqn3}
\end{equation}
Where $\lambda$ and $\mu_0$ are the relaxation time and zero-shear viscosity of the liquid. Given the importance of the bag breakup mode, experiments are performed at a fixed $We$ value of $\sim$12.5, where bag breakup mode is observed for all the test liquids. The $El$ is varied in the range of $\sim10^{-2}-10^{-4}$. Some of these polymeric droplets, investigated in the present work, have low enough elasticity to show fragmentation during the observation timescale of a bag breakup experiment. This is in contrast with the work of \citet{theofanous2013physics}, which reports that the fragmentation of a polymeric droplet is not possible in the bag breakup regime or, to say, RTP regime in the author's way of classification.

\section{Materials and methods}
\label{sec:matrials_methods}

\subsection{Experimental setup}

\begin{figure*}
\includegraphics[width=0.9\linewidth]{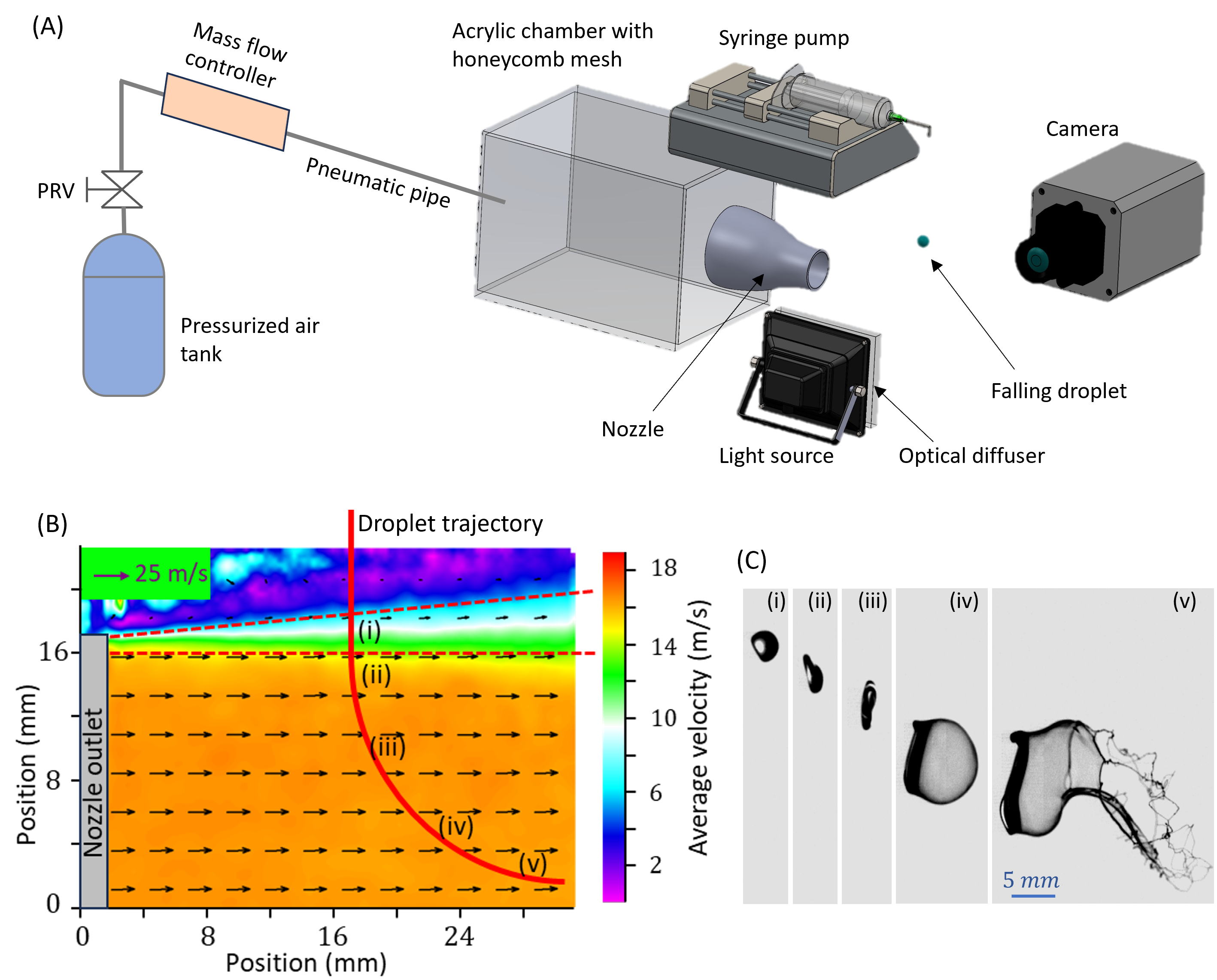}
\centering
\caption{(A) Schematic representation of the experimental setup. (B) Time-averaged airflow velocity field at nozzle exit measured using the PIV technique. The velocity field is shown for the region above the nozzle center line, which can be assumed to be symmetric for the lower half. (C) Snapshots of droplet deformation and breakup at different locations marked in (B).}
\label{fig:exp_setup}
\end{figure*}

Experiments are performed by allowing the liquid droplet to fall in a horizontal air stream coming out from a nozzle (outlet diameter= 34 mm) as shown in Figure \ref{fig:exp_setup}A. The air nozzle is fitted tightly on a cuboidal acrylic chamber (20 cm$\times$20 cm$\times$38 cm) equipped with two layers of honeycomb mesh to laminarize the airflow at the nozzle inlet. The constant air supply is provided by a large pressurized air reservoir (capacity 10 $m^3$), which is connected to the acrylic chamber with the help of a pneumatic pipe. A pressure regulating valve (PRV) and an electronically-controlled mass flow controller (Alicat scientific, 3000 SLPM) are installed in this pneumatic pipe. The PRV steps down the pressure between the air tank and the experimental section, while the mass flow controller allows precise airflow rate control through the nozzle.

The liquid droplet is generated by slowly pumping ($\sim$40 $\mu$L/min) the liquid out through a $90^o$ bent Nordson needle connected to a syringe pump. This syringe pump is installed on a vertical rail so that the falling height (in turn, velocity) of the liquid droplet can be adjusted suitably. The liquid droplet is dispensed at an axial distance of $\sim$15 mm from the nozzle outlet. A high-speed camera (Phantom SA-5) with its optical axis perpendicular to both the airflow and the droplet falling direction is installed to record the shadowgraphic images of the breakup phenomenon. The light source (Veritas Constellation 120E, pulse width 2 $\mu$s) of the shadowgraphic imaging setup is operated in strobe mode and in synchronization with the high-speed camera to reduce the streaking of images due to motion blur. Images are recorded at 20000 frames per second with an exposure time of 50 $\mu$s and an observation window of 1024$\times$368 pixels. A spatial resolution of $\sim$103 $\mu$m/pixel is achieved by connecting a macro lens (Sigma DG macro HSM) with a fixed focal length of 105 mm to the high-speed camera.

Particle image velocimetry (PIV) experiments are performed separately on a High-speed PIV system (LaVison) by seeding the acrylic chamber with DEHS droplets to characterize the airflow coming out of the nozzle. The PIV data is analyzed using commercial software, LaVision Davis 8.4. Time-averaged air velocity field obtained from PIV measurements for airflow rate, $Q=$800 SLPM, is shown in Figure \ref{fig:exp_setup}B. The velocity field is shown for the upper half from the nozzle centerline, and it can be assumed symmetric for the lower half. It is observed that airflow follows a plug flow-like profile in the central core region, except for a thin shear layer around the air jet indicated by red dashed lines in Figure\ref{fig:exp_setup}B. By adjusting the droplet falling height, it is ensured that the falling droplet spends minimum time in the shear layer, and major breakup events happen in the region of uniform airflow. An approximate trajectory of the falling droplet is shown overlapped with the airflow field in Figure \ref{fig:exp_setup}B (Solid red curve). Figure \ref{fig:exp_setup}C shows the snapshots of the droplet breakup phenomenon at different positions in the trajectory numbered (i)-(v) in  Figure \ref{fig:exp_setup}B. From the PIV measurement of the airflow, it is estimated that the shear layer has a thickness of $\sim$2 mm at the axial position where the droplet enters the air stream. This thickness is smaller than the average droplet diameter ($\sim$3.5 mm), which further ensures that major events of droplet breakup happen in the uniform airflow region. Since the airflow profile is similar to plug flow, the air velocity, $U_g$ for Weber number calculation, is estimated based on the average velocity ($=Q/A$), where $A$ is the cross-section area of the nozzle exit. The calculated value of the average air velocity closely matches the PIV measurement in the region of uniform airflow.

\subsection{Liquid properties}

\begin{table}
  \begin{center}
\def~{\hphantom{0}}
  \begin{tabular}{l|c|c|c|c|c|c|c}
Name / & c & $c/c^*$ & $\mu_0$ & $\lambda$ & $\gamma$ & $El$ & $Oh$ \\
 Polymer type & (\% w/w) &  & (mPa-s) & (ms) & (mN/m) & &  \\[6pt]
 \hline

Solvent & 0 & 0 & 3.3 & 0 & 68 & 0 & 6$\times 10^{-3}$ \\
PEO-4M & 0.15 & 2.14 & 17 & 47.6 & 60 & 6$\times 10^{-2}$ & 4$\times 10^{-2}$ \\
PEO-4M & 0.016 & 0.23 & 4.3 & 13.3 & 60 & 4$\times 10^{-3}$ & 1$\times 10^{-2}$ \\
PEO-4M & 0.008 & 0.11 & 4.1 & 4.6 & 60 & 1.4$\times 10^{-3}$ & 1$\times 10^{-2}$ \\
PEO-4M & 0.003 & 0.04 & 3.4 & 3.3 & 66 & 8.5$\times 10^{-4}$ & 7$\times 10^{-3}$ \\
PEO-0.6M & 0.6 & 2.61 & 22 & 5.7 & 60 & 9.2$\times 10^{-3}$ & 5$\times 10^{-2}$ \\
PEO-0.6M & 0.3 & 1.30 & 10 & 3.4 & 60 & 2.5$\times 10^{-3}$ & 2$\times 10^{-2}$ \\
PEO-0.6M & 0.1 & 0.43 & 5.4 & 1.8 & 60 & 7$\times 10^{-4}$ & 1$\times 10^{-2}$ \\
PEO-0.6M & 0.03 & 0.14 & 3.6 & 0.8 & 60 & 2.2$\times 10^{-4}$ & 7$\times 10^{-3}$ \\

  \end{tabular}
  \caption{Properties of the test liquids.}
  \label{liquid_properties}
  \end{center}
\end{table}

\begin{figure*}
\includegraphics[width=0.9\linewidth]{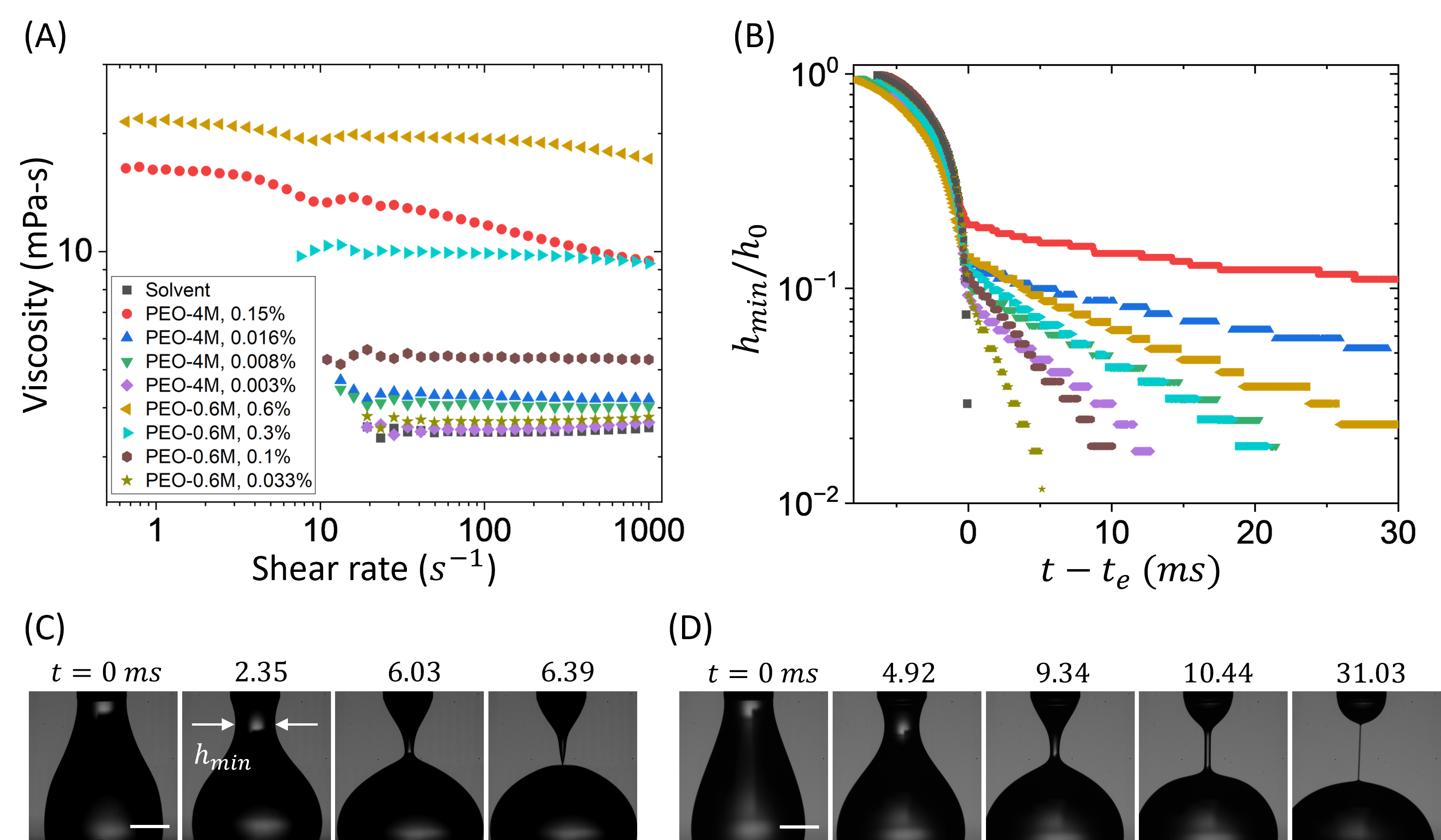}
\centering
\caption{Shear and elongational rheology of all the liquids investigated in the present work. (A) Variation of apparent shear viscosity with shear rate. (B) Transient variation of minimum neck diameter ($h_{min}$) for the liquid bridge obtained from the CaBER-DoS experiment. The symbols and color code used in (B) are the same as in (A). Snapshots of the CaBER-DoS experiments for (C) Newtonian solvent and (D) polymeric liquid PEO-0.6M, $c=0.6\%$. Scale bar in (C) and (D) represents 1 mm.}
\label{fig:rheology} 
\end{figure*}

We performed bag breakup experiments with a Newtonian solvent and eight different polymeric solutions. The Newtonian solvent is a solution of water and glycerol with 60 and 40 as their respective weight percentage. Polymeric solutions are obtained by dissolving small amounts of polyethylene oxide (PEO) in the Newtonian solvent. PEO with two different viscosity-averaged molecular weights ($M_w$), $4\times10^6$ g/mol referred to as PEO-4M and $6\times10^5$ g/mol referred to as PEO-0.6M, purchased from Sigma Aldrich is used. The concentration, $c$, and other properties of liquids investigated in the present work are provided in Table \ref{liquid_properties}. The critical overlap concentration, $c^*$ for PEO-4M and PEO-0.6M, are 0.07$\%$ and 0.23$\% (w/w)$, respectively. This is calculated using the Flory relation for flexible polymer solution, $c^*=1/[\eta]$, where $[\eta]$ is the intrinsic viscosity of the polymeric solution estimated using the composite Mark-Houwink-Sakurada equation $[\eta]=0.072\times (M_w)^{0.65}$ \citep{tirtaatmadja2006drop}. The surface tension, $\gamma$, is obtained by capturing pendant drop images and then analyzing them using the Pendent drop plugin (version 2.0.1) available as an open-source tool in ImageJ software. An approximate average value of $\gamma$ obtained from three different trials is reported in Table \ref{liquid_properties}. The shear rheology of test liquids is performed at a fixed temperature of 25 $^o$C using a cone and plate geometry (plate diameter 40 mm, cone angle 1$^o$) of a commercial rheometer (Anton Paar, MCR 702). The apparent shear viscosity curve for all the test liquids is presented in Figure \ref{fig:rheology}A. Generally the high molecular weight PEO solutions shows shear-thinning behavior if the solvent is plain DI water. However, the degree of shear-thinning can be reduced by choosing a solvent with higher viscosity, as done in the present work by selecting water-glycerol-based solvent \citep{mun1999atomisation,keshavarz2016ligament}. The polymeric liquids investigated in the present work exhibit Boger fluid-like behavior, i.e., elastic fluids with constant shear viscosity. This makes it straightforward to calculate the $Oh$ based on the zero-shear viscosity, $\mu_{0}$ as reported in Table \ref{liquid_properties}. Since $Oh$ is less than 0.1 for all the polymeric droplets, the effect of liquid shear-viscosity on the droplet breakup can be neglected, and due to Boger fluid-like behavior, the effect of shear-thinning nature is eliminated. Hence, any deviation from the Newtonian behavior can be attributed to the liquid elasticity alone. 

The relaxation time $\lambda$ of the polymeric solutions is obtained by performing custom-made capillary breakup elongation rheometry by dripping-onto-substrate (CaBER-DoS), following the protocol available in literature \citep{dinic2017pinch,chandra2021contact}. In CaBER-DoS rheometry, a drop from a needle is deposited gently on a clean glass substrate, and the minimum neck diameter, $h_{min}$ of the liquid bridge formed between the needle and the substrate, is tracked with time. Figure \ref{fig:rheology}C and \ref{fig:rheology}D shows the snapshots of the CaBER-DoS experiments for the Newtonian solvent and a polymeric solution (PEO-0.6M, 0.6$\%$). The transient variation of minimum neck diameter normalized by the needle diameter, $h_0\sim1.27$ mm for all the test liquids, is presented in Figure \ref{fig:rheology}B. It can be seen that the thinning of the polymeric liquid bridge is similar to the Newtonian solvent at an early stage ($t\leq t_e$). This is known as the inertia-capillary (IC) regime, where liquid inertia balances the capillary force, and the minimum neck diameter follows a power-law thinning with time such that $\frac{h_{min}}{h_0}\propto (t_e-t)^{2/3}$. However, at some instant, $t=t_e$, which depends upon the polymer type and concentration, viscoelastic stress becomes dominant, and now the liquid bridge thins exponentially with time. This is called the elasto-capillary (EC) regime and its occurrence is attributed to the transition of polymer molecules from a coiled to a stretched configuration leading to high viscoelastic stresses. Elongational relaxation time of the polymeric liquid is obtained by fitting an exponential curve in the EC regime, such that $\frac{h_{min}}{h_0}=Ae^{-B(t-t_e)}$ and $\lambda=\frac{1}{3B}$.

\section{Results and discussion}
\label{sec:results_discussion}

\subsection{General description of the bag breakup mode}
\label{subsec:genral_description}


\begin{figure*}
\includegraphics[width=1\linewidth]{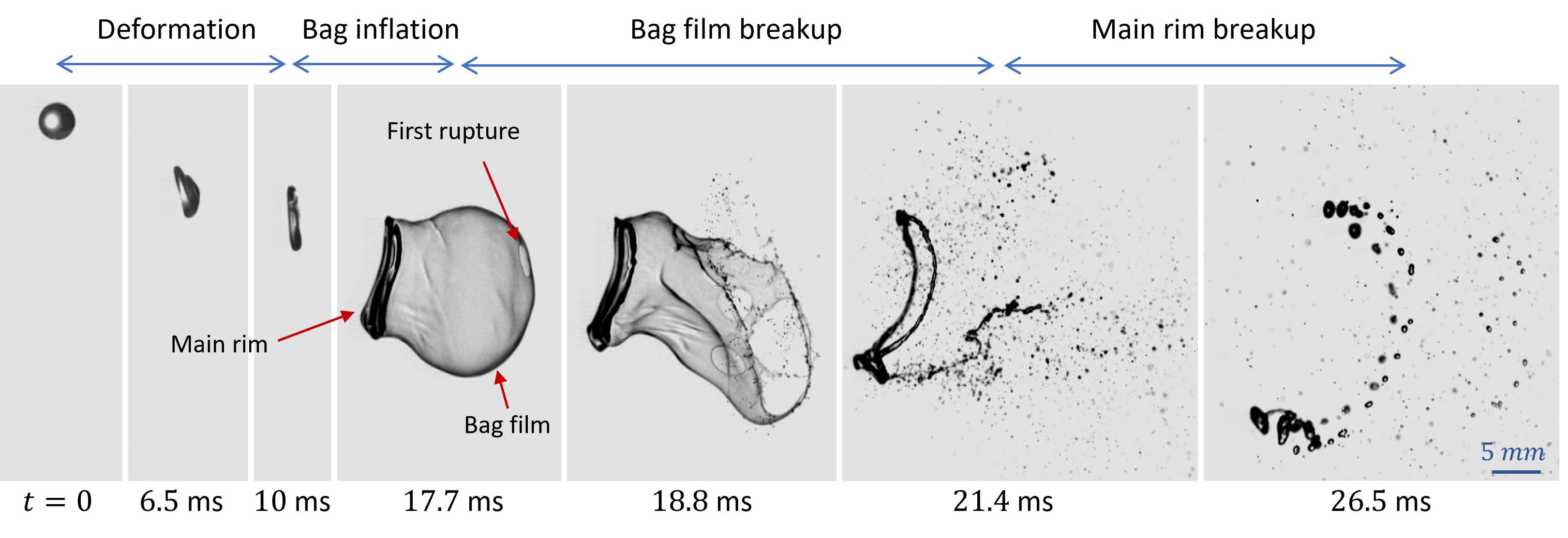}
\centering
\caption{Different stages in a bag breakup mode.}
\label{fig:general_bag} 
\end{figure*}

Different stages and terminologies associated with the aerodynamic bag breakup of a droplet are shown in Figure \ref{fig:general_bag}, with the example of a Newtonian solvent droplet undergoing bag breakup at $We=$12.5. The breakup starts at the time, $t=0$ when a spherical droplet of diameter $D_0$ enters the continuous air stream. The first stage is the deformation stage, during which the spherical droplet transforms into a thin disk-like shape. In some literature, this is also referred to as the breakup initiation phase \citep{jackiw2021aerodynamic,flock2012experimental}. The end of the deformation stage is marked by the moment at which the deforming liquid disk achieves its minimum thickness (Figure \ref{fig:general_bag}, $t=$ 10 ms). This instant is known as the initiation time and is generally denoted by $t_i$. The next stage is the bag inflation stage, during which, the air blows the liquid disk in the form of a bag such that the central part of the disk forms a thin bag film and the periphery of the disk forms a thicker main rim attached to the bag (Figure \ref{fig:general_bag}, $t=$ 17.7 ms). As the bag size grows, the bag film becomes thinner, and at some instant, a rupture appears on the bag film, which marks the end of the bag inflation stage (Figure \ref{fig:general_bag}, $t=$ 17.7 ms). Multiple holes form and grow on the bag film as the hole, due to initial rupture, expands. This is termed as the bag film breakup stage. Here, the bag film undergoes complicated mechanisms like hole-rim destabilization and hole-hole merging phenomenon (discussed in detail later), which result in the fragmentation of the bag film into fine daughter droplets in the case of Newtonian liquids, although not necessarily for polymeric droplets. The end of the bag film breakup stage is identified by the instant when the sheet-like structure of the bag film has completely disappeared (Figure \ref{fig:general_bag}, $t=$ 21.4 ms). The final stage in bag breakup is the main rim breakup stage, where the toroid-shaped main rim breaks up, resulting in larger fragments compared to fragments obtained from the bag film breakup (Figure \ref{fig:general_bag}, $t=$ 26.5 ms). Considering the example presented in Figure \ref{fig:general_bag}, the relative time spent in different stages can be estimated. For this particular example, the time spent in deformation, bag inflation, bag film breakup, and main rim breakup stage is approximately 38$\%$, 29$\%$, 14$\%$, and 19$\%$, respectively of the total breakup time. It should be noted that the bag film breakup is the first stage where actual fragmentation occurs. Consequently, the primary focus of the present work is to comprehensively investigate the bag film breakup process and, hence, establish the criterion for fragmentation of a polymeric droplet.

\subsection{Droplet deformation and bag inflation}
\label{subsec:deformation_inflation}

\begin{figure*}
\includegraphics[width=0.9\linewidth]{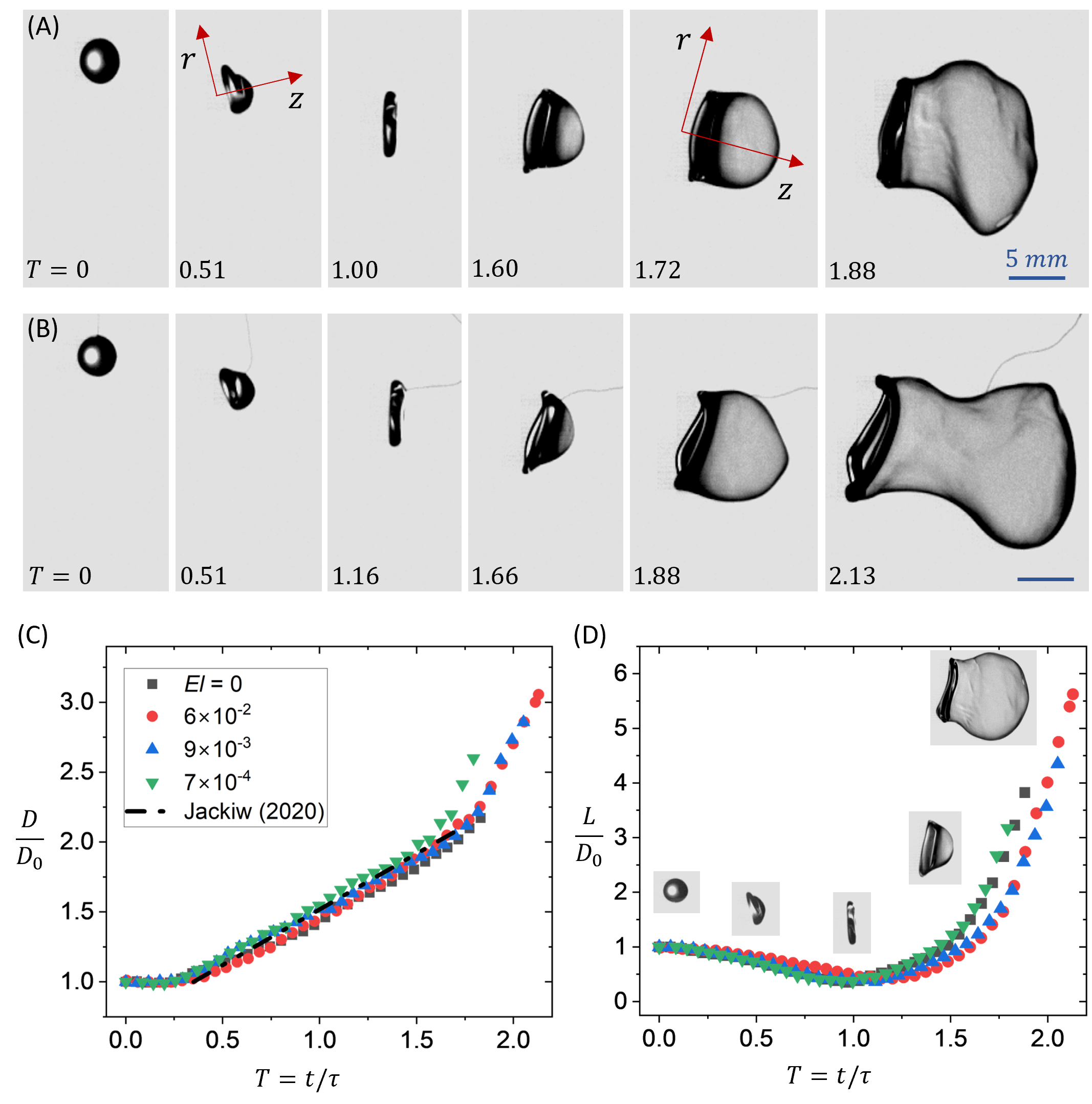}
\centering
\caption{Experimental images showing deformation and bag inflation at $We\approx12.5$ for a droplet with (A) $El=0$ and (B) $El=6\times10^{-2}$. The extent of liquid mass, measured in the (C) $r-$direction and (D) $z-$direction, till the instant of the first rupture. Symbols and color code used in (D) are the same as in (C).}
\label{fig:deformation_inflation} 
\end{figure*}

Droplet deformation and bag inflation set the stage for subsequent fragmentation in the bag breakup mode. Figure \ref{fig:deformation_inflation}A and \ref{fig:deformation_inflation}B shows the experimental snapshots during the droplet deformation and the bag inflation stage for the Newtonian solvent and a polymeric droplet with $El=6\times10^{-2}$, which is the highest $El$ investigated in the present work. Here, temporal change is represented in terms of non-dimensionalized time, $T=\frac{t}{t_I}$ such that, $t_I=\frac{D_0}{U_g}\sqrt{\frac{\rho_l}{\rho_g}}$ represent the inertial timescale \citep{nicholls1969aerodynamic}. Droplet deformation is the first stage in the breakup process which happens when the falling droplet enters the air stream, and the configuration becomes similar to the uniform gas flow over a deformable spherical body. The air stagnation pressure at the windward side of the droplet deforms it from a spherical to a flat disk shape. It should be noted that since the droplet enters through a diverging shear layer, the plane of the windward disk on the droplet surface is not perpendicular to the main airflow direction, and the tilt angle changes with time \citep{xu2023transitions}. Therefore it is convenient to describe the deformation process in a reference frame attached with the translating and deforming liquid mass as shown in Figure \ref{fig:deformation_inflation}A ($T=0.51$ and $T=1.72$). The air pressure distribution along the periphery of the liquid droplet causes the liquid to flow from the windward pole to the equator of the droplet, i.e., along the $r-$direction \citep{jackiw2021aerodynamic}. Ultimately transforming the liquid mass into the shape of a disk. In side-view imaging, only the thickness of the disk can be visualized, as shown at $T=1.00$ in Figure \ref{fig:deformation_inflation}A and $T=1.16$ in Figure \ref{fig:deformation_inflation}B. The time instant at which the liquid disk attains its minimum thickness is termed the breakup initiation time and is denoted by $T_i$ in non-dimensionalized form. Immediately after this instant, the periphery of the disk pinches in the form of a toroidal rim, maintaining its constant minimum thickness. While the central part of the disk undergoes further thinning, still connected to the toroidal rim, making it susceptible to be blown downstream by the airflow and hence commencing the bag inflation process \citep{jackiw2021aerodynamic}. In the case of polymeric droplets with sufficiently high elasticity, generally, a liquid tail is formed when a falling drop is generated from a needle which can be seen attached on top of the liquid droplet in Figure \ref{fig:deformation_inflation}B. This is unavoidable in working with polymeric droplets; however, they don't significantly affect the droplet breakup dynamics \citep{theofanous2013physics}.

The droplet deformation and bag inflation are quantified in terms of $D$ and $L$, representing the maximum extent of the liquid mass along the $r$ and the $z-$directions. The temporal evolution of $D$ and $L$ for the Newtonian solvent and three different polymeric droplets are presented in Figure \ref{fig:deformation_inflation}C and \ref{fig:deformation_inflation}D. Representative experimental images from the Newtonian case are shown as inset in Figure \ref{fig:deformation_inflation}D. It should be noted that for $T\leq T_i$, $D$ and $L$ represent the diameter and thickness of the liquid disk formed during the deformation stage. While, for $T>T_i$, they represent the main rim diameter, and the bag size along $z-$ direction, respectively. From Figure \ref{fig:deformation_inflation}C and \ref{fig:deformation_inflation}D, it can be observed that even with three orders of magnitude increase in the $El$, the presence of liquid elasticity does not play any significant role during the deformation stage. However, the role of liquid elasticity can be observed in the bag inflation stage in terms of larger bag size before the first rupture. The insignificant role of liquid elasticity during the deformation stage ($T\leq T_{i}$) can be explained following a philosophy similar to the observation of capillary pinch-off experiment for CaBER-DoS rheometry(Figure \ref{fig:rheology}B). Where a polymeric liquid shows similar behavior to that of the Newtonian solvent at an early stage (IC regime). However, it shows a drastic deviation from the Newtonian behavior at a later stage (EC regime). The transition from IC to EC regime happens when the elongational strain rate ($\dot{\epsilon}$) in the polymeric liquid reaches a critical value, $\dot{\epsilon_c}$, such that extensional stress on a polymer molecule due to the background flow of Newtonian solvent is strong enough to cause the coil-stretch transition of the polymer molecule \citep{wagner2015analytic,rajesh2022transition}. Consequently, if $\dot{\epsilon}<\dot{\epsilon_c}$ for a given process involving polymeric liquid, then it is expected that the observations will be similar to that of a Newtonian liquid and the elastic properties will not play a major role. This can be checked for the present case by having the knowledge of $\dot{\epsilon}$ relevant to the droplet deformation process and $\dot{\epsilon_c}$ corresponding to different polymeric solutions tested in the present work. \citet{rajesh2022transition} performed capillary pinch-off experiments with various combinations of water-glycerol-PEO solution and provided the following power-law fit (Figure 9 of the mentioned reference) to estimate the critical extensional strain rate.
\begin{equation}
\dot{\epsilon_c}=0.1 + \left( \frac{2}{3\lambda} \right)^{0.45}
\label{eqn4}
\end{equation}
Here, the units of $\dot{\epsilon_c}$ is in ms$^{-1}$ and $\lambda$ is in ms. Among all the polymeric liquids tested in the present work, the minimum $\dot{\epsilon_c}$ is $\approx$ 246 s$^{-1}$ which corresponds to polymeric liquid with the highest $\lambda=$ 47.6 ms. The relevant strain rate in the liquid phase during droplet deformation can be estimated by having knowledge of the liquid flow field, which is generally idealized as a bi-axial extensional flow \citep{jackiw2021aerodynamic,kulkarni2014bag,villermaux2009single}. Recently, \citet{jackiw2021aerodynamic} provided a model to predict the constant rate of radial expansion, $\dot{R}$, of the deforming droplet by balancing dynamic liquid pressure with the outside air pressure and Laplace pressure jump.
\begin{equation}
\frac{\dot{R}}{D_0/2}=\frac{1}{\tau_I} \left ( \frac{\alpha}{2} \right )^2 \left (1-\frac{128}{\alpha^2 We} \right ) T_{bal}
\label{eqn5}
\end{equation}

Here, $\alpha$ is the stretching rate of airflow which has a value of 6 and $\pi$/4 for the airflow over a sphere and disk, respectively. Since the constant radial expansion rate is decided early in the deformation process, when the liquid mass has more sphere-like shape, therefore, the value of $\alpha$ can be taken as 6. $T_{bal}$ is the initial flow balancing time period and its most suitable is $\approx1/8$ estimated from the experimental observation \citep{jackiw2021aerodynamic}. The deformation rate predicted from Equation \ref{eqn5} is in good agreement with the present experiments, as shown in Figure \ref{fig:deformation_inflation}C. The extensional strain rate in the liquid phase during deformation can be calculated as $\dot{\epsilon}=2\dot{R}/R$. Since $\dot{R}$ is constant and $R$ increases with time, the value of $\dot{\epsilon}$ will decrease with time. The maximum value of $\dot{\epsilon}$ will ideally occur at $T=0$, estimated as $\approx$ 200 s$^{-1}$ for the present experiments, which is less than the minimum $\dot{\epsilon_c}$ required considering all the polymeric liquids investigated in the present work. This explains our observation of liquid elasticity playing an insignificant role during the deformation stage. One should note that, during the inflation stage, bag film thins independently from the main rim. Therefore, the $\dot{\epsilon}$ in the bag film will be given by $\dot{\delta}/\delta$ where $\delta$ is the thickness of the bag film. Since $\delta$ is very small ($\sim10^0 \mu$m), the $\dot{\epsilon}$ in the bag film can become sufficiently high to overcome $\dot{\epsilon_c}$ and hence allowing the liquid elasticity to play a dominant role. In such a scenario, it is expected that liquid elasticity will try to homogenize the sheet thickness throughout the bag film, similar to capillary breakup, where the effect of liquid elasticity is observed as a uniform radius along the length of the liquid bridge (Figure \ref{fig:rheology}D). This hypothesis has the potential to explain the large bag before the first rupture observed with polymeric liquids. However, the thickness modulation in the bag film is difficult to verify experimentally, and a detailed numerical simulation will be required to verify this hypothesis further.

\subsection{Bag film breakup}
\label{subsec:bag_film_breakup}

\begin{figure*}
\includegraphics[width=0.9\linewidth]{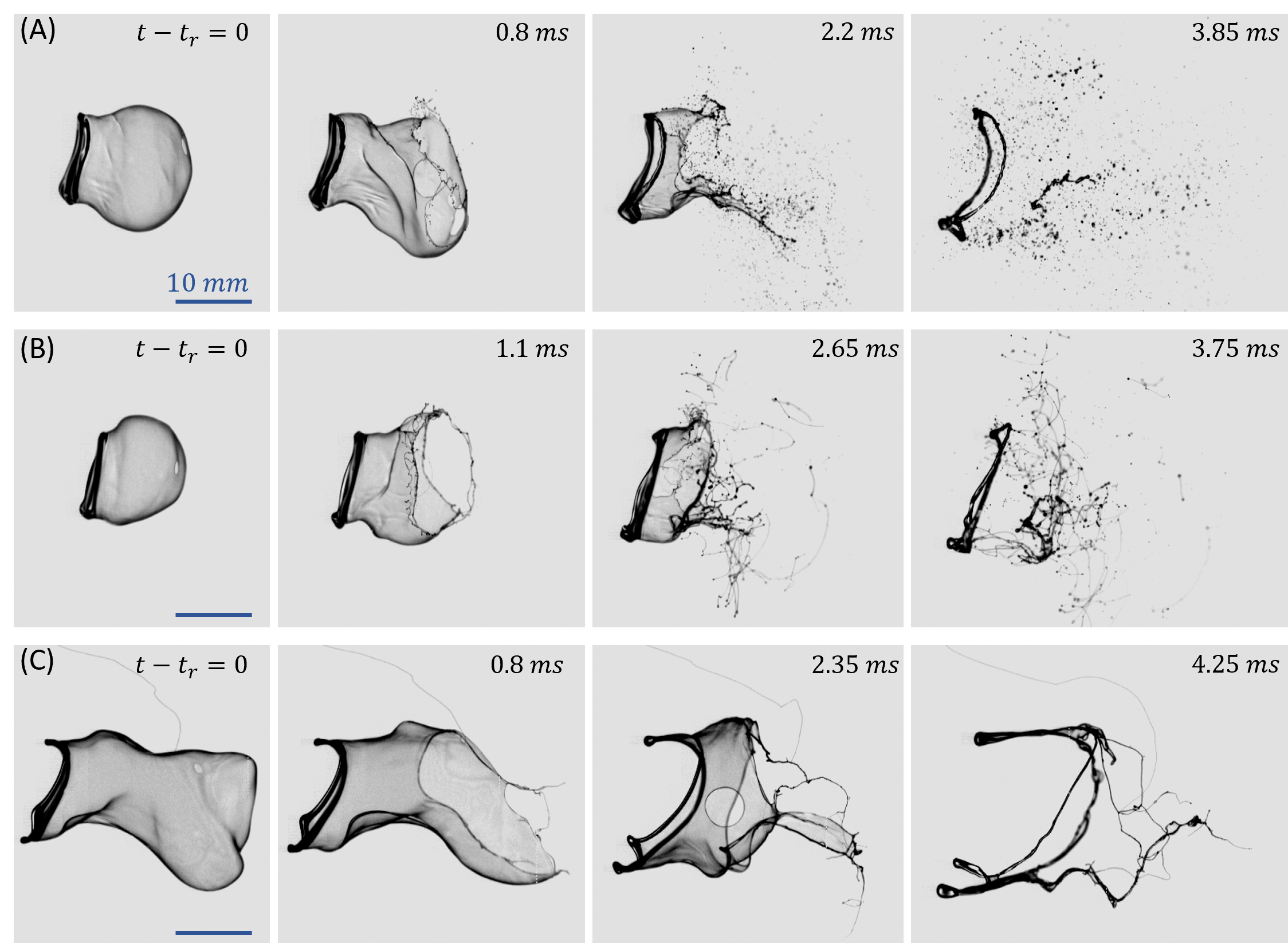}
\centering
\caption{Bag film breakup for (A) the Newtonian solvent droplet ($El=0$), and polymeric droplets with (B) $El=8\times10^{-4}$, (C) $El=6\times10^{-2}$. }
\label{fig:bag_film_breakup} 
\end{figure*}

The breakup of bag film is the most important aspect of the bag breakup mode because it is the first stage where actual fragmentation happens, and the fragments obtained from this stage decide the lower limit of fragment size distribution. Figure \ref{fig:bag_film_breakup} shows the bag film breakup, starting from the instant of first rupture till the sheet-like structure of the bag film is completely exhausted, for the Newtonian solvent and polymeric droplet with two different $El$. The study of this stage is difficult because of the multitude of lengthscales and timescales involved in the process. Typically the bag size is more than a few millimeters, while the film thickness and fragments obtained from the bag film breakup are micron-sized. Further, the breakup of the bag film happens in a few milliseconds while the liquid mass is convecting downstream of the airflow. Adding up all the complexity, it requires imaging with high temporal and spatial resolution along with a large field of view, which are difficult to accommodate together in a high-speed camera. Since the objective of the present work is to identify the occurrence of fragmentation, but not the fragment size measurement, therefore we compromise with the spatial resolution. Another challenge in this stage is that the cause of hole initiation in the bag film is not exactly known, with possible origins from- thermal fluctuations, disjoint pressure, RT instability due to airflow, and the Marangoni flows in the liquid phase due to the presence of contaminants \citep{neel2018spontaneous,vledouts2016explosive}. However, previous studies have shown that once a hole is formed in the bag film, the further hole growth and subsequent fragmentation is governed only by the liquid properties, and the effect of airflow can be neglected, at least in the bag breakup regime of millimeter-sized Newtonian droplets \citep{chou1998temporal,jackiw2022prediction}. Considering the arguments provided in these studies, we can assume a similar behavior for the bag breakup of a polymeric droplet.

Fragmentation of the bag film proceeds through intermediate ligament formation, which is formed through two different mechanisms. First, through the destabilization of the rim bounding a hole in the bag film. Second, due to the merging of two expanding holes in the bag film. These liquid ligaments formed out of the bag film further undergo capillary pinch-off to finally result in daughter fragments. It can be observed from Figure \ref{fig:bag_film_breakup}A that the bag film of the Newtonian solvent is completely fragmented into smaller droplets (see supplementary movie 1). While the polymeric droplets are either partially fragmented (Figure \ref{fig:bag_film_breakup}B) or there is no actual fragmentation, and the liquid mass remains in the form of an interconnected web of ligaments (Figure \ref{fig:bag_film_breakup}C), depending upon the liquid elasticity (see supplementary movie 2 and 3). Given a sufficiently long time, liquid ligaments, even with very high elasticity, will break into smaller fragments. However, the time provided for fragmentation cannot be infinitely long. In the case of practical applications, the time available for fragmentation is limited by the distance between the spray source and the target. Whereas in laboratory scale experiments, this time is limited either by the residence time of the liquid mass in the air stream or the observation window of the high-speed imaging system. Therefore, it is logical to talk about a criterion deciding the occurrence of fragmentation only in a specified finite time window. The absence of fragmentation can be prescribed in the following form.
\begin{equation}
\frac{t_{cp}}{t_{f}}>1
\label{eqn6}
\end{equation}
Here, $t_{cp}$ is the capillary pinch-off time for the ligaments formed out of the bag film, and $t_f$ is the time of flight or the observation time span after the formation of ligaments. Apart from liquid properties mentioned in Table \ref{liquid_properties}, the value of $t_{cp}$ will depend upon the ligament diameter, $d_l$. The estimation of $d_l$ requires knowledge of the mechanism leading to ligament formation, which is discussed in the next two subsections (\ref{subsec:rim_destabilization} and \ref{subsec:hole_merging}).

\subsection{Hole rim destabilization}
\label{subsec:rim_destabilization}

\begin{figure*}
\includegraphics[width=0.85\linewidth]{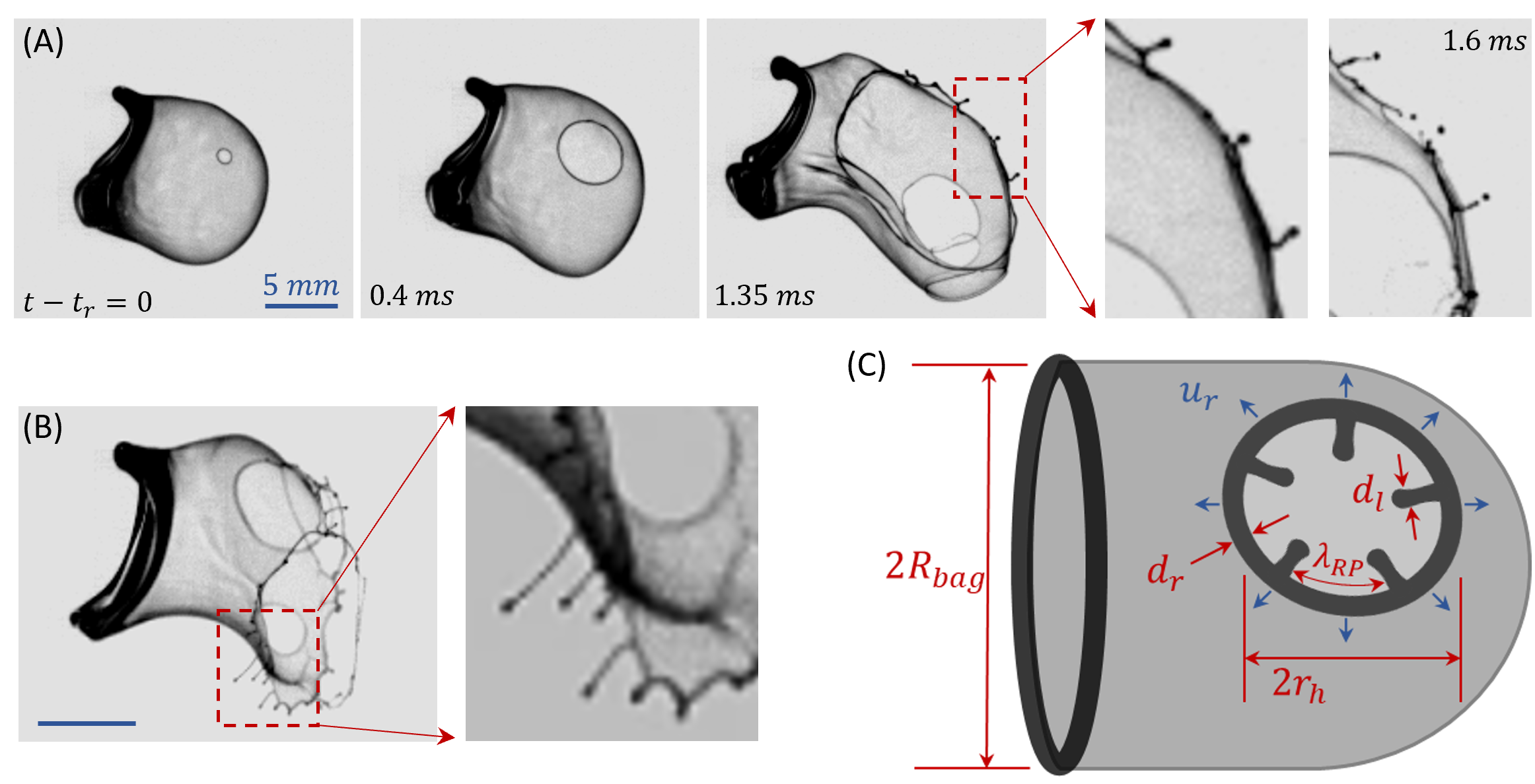}
\centering
\caption{Ligament formation due to destabilization of a rim receding on the bag film for (A) the Newtonian solvent droplet ($El=0$) and (B) a polymeric droplet with $El=8\times10^{-4}$. (C) Schematic of the rim destabilization showing different variables involved in the process.}
\label{fig:rim_destabilization} 
\end{figure*}

The first mechanism leading to ligament formation is the destabilization of a rim receding on the bag film, as shown in Figure \ref{fig:rim_destabilization}A. A hole formed on the bag film is unstable due to the unbalanced surface tension force on the liquid edge bounding the hole. At this point, any elastic energy stored in the liquid phase during the bag inflation stage also supports the hole-opening process \citep{di2022bubble} (See Appendix \ref{appA}). As the hole radius grows with time, the corresponding liquid volume from the bag film gets collected in the form of a rim around the hole. This rim, receding on the bag film, is susceptible to RP instability which appears as corrugations on the rim and subsequently grows as dangling ligaments connected to the rim \citep{wang2018universal,jackiw2022prediction}, as shown for the Newtonian solvent in Figure \ref{fig:rim_destabilization}A ($t-t_r$=1.35 ms). Another example of ligament formation by rim destabilization for a polymeric droplet with $El=8\times10^{-4}$ is shown in Figure \ref{fig:rim_destabilization}B. A schematic representation of ligaments formed due to hole rim destabilization, along with relevant physical variables involved in the process, is shown in Figure \ref{fig:rim_destabilization}C. In such a rim destabilization process, the wavelength of corrugations, $\lambda_{RP}$ is set by the RP instability, whereas the rim thickness, $d_r$ at the time of destabilization is decided by the acceleration of the rim. \citet{wang2018universal} showed that the rim destabilization happens when the local instantaneous Bond number, $Bo$, reaches unity.
\begin{equation}
Bo = \frac{\rho_l a_r d_r^2}{\gamma} = 1
\label{eqn7}
\end{equation}
Here, $a_r$ is the acceleration experienced by the receding rim. The criteria mentioned in Equation \ref{eqn7} applies to both the Newtonian and the viscoelastic liquids but only up to a certain degree of liquid elasticity. Similarly, in the present work also, we observed the rim destabilization phenomenon only for the polymeric droplet with low $El$ ($\leq 1.4\times10^{-3}$) and not for the higher ones. \citet{wang2018universal} provided the upper limit of liquid elasticity for applicability of Equation \ref{eqn7} in terms of a phenomenological Deborah number which cannot be directly applied to the present experiments. However, the relaxation time of present polymeric liquids showing rim destabilization is small compared to the ones reported by \citet{wang2018universal}. Therefore, we consider that the criterion of $Bo=1$ is applicable for the present polymeric liquids, which show rim destabilization. It can be assumed that the diameter of the ligaments formed from the rim will have a similar magnitude as the rim thickness at the time of destabilization, i.e., $d_l \approx d_r$. Equation \ref{eqn7} can be used to calculate $d_r$ if $a_r$ is known. Similar to a static bubble bursting \citep{lhuissier2012bursting},  $a_r$ can be estimated from the centripetal acceleration experienced by the rim due to its recession over a curved surface (bag film).
\begin{equation}
a_r = \frac{u_r^2}{R_{bag}}
\label{eqn8}
\end{equation}
Here, $u_r=\dot{r_h}$ is the receding velocity of the rim such that $r_h$ represents the radius of the hole (see Figure \ref{fig:rim_destabilization}C). $R_{bag}$ is the radius of the curved surface approximated as the radius of the main rim at the time of the first rupture in the bag film. For low-viscosity Newtonian liquids, $u_r$ is given by the Taylor-Culik relation.
\begin{equation}
u_r = \sqrt{\frac{2\gamma}{\rho_l \delta}}
\label{eqn9}
\end{equation}
In the case of viscoelastic liquids, the value of $u_r$ can be higher than the value predicted from the Taylor-Culik relation due to the release of the elastic energy stored during the bag inflation process. However, in the present case, the contribution of liquid elasticity is negligible (see Appendix \ref{appA}). Therefore, Equation \ref{eqn9} can be applied to all the liquids investigated in the present work. Equations \ref{eqn7}, \ref{eqn8}, and \ref{eqn9} can be re-arranged to estimate the ligament diameter.
\begin{equation}
d_l \approx \sqrt{\delta R_{bag}/2}
\label{eqn10}
\end{equation}

\begin{figure*}
\includegraphics[width=0.85\linewidth]{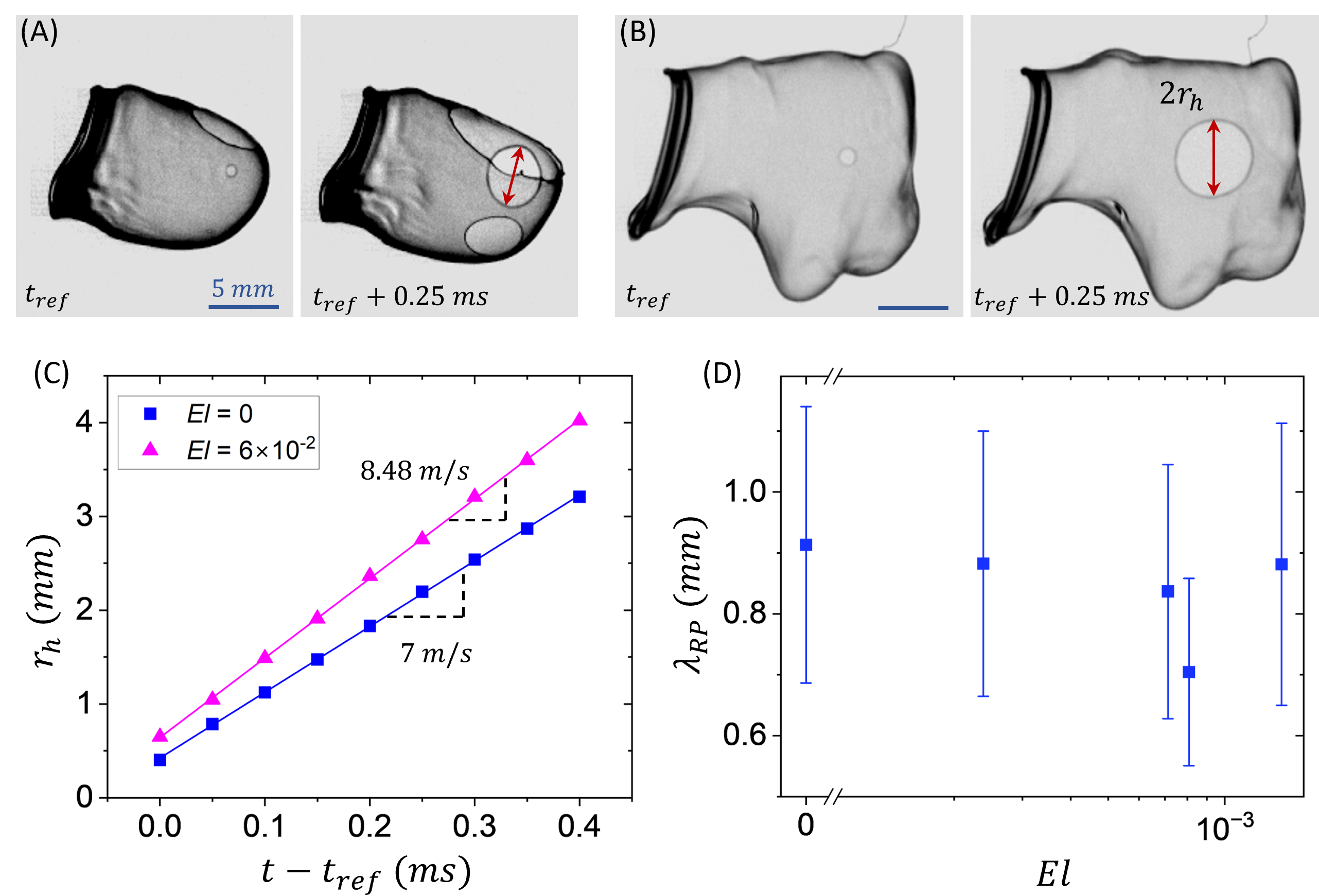}
\centering
\caption{Estimation of the bag film thickness by measuring the hole-opening velocity on the bag film. Experimental images of the hole-opening process on the bag film of (A)  Newtonian solvent, and (B) a polymeric droplet with $El=6\times10^{-2}$. (C) Transient variation of the hole radius for the cases shown in (A) and (B). (D) Spacing between ligaments formed due to receding rim destabilization.}
\label{fig:hole_opening} 
\end{figure*}
It is evident from Equation \ref{eqn10} that, in order to estimate $d_l$, the value of $R_{bag}$ and $\delta$ is required. The former can be directly measured from the experimental images, while the latter is difficult to measure directly. However, $u_r$ can be easily measured from the experiments, and then Equation \ref{eqn9} can be used to get an estimate of $\delta$. To measure $u_r$, the radius, $r_h$, of a hole formed on the bag film, is tracked with time. For this purpose, the holes that are properly visible in the side-view imaging are selected, as shown in Figure \ref{fig:hole_opening}A and \ref{fig:hole_opening}B with the corresponding temporal data plotted in Figure \ref{fig:hole_opening}C. Here $t_{ref}$ is a reference time instant at which the hole under observation becomes visible on the bag film. A linear fit matches well with the temporal evolution of $r_h$, and the slope of this linear fit is considered as the value of $u_r$. It can be observed that the value of $u_r$ is slightly higher for polymeric liquid ($\approx$ 8.48 m/s) compared to the Newtonian solvent ($\approx$ 7 m/s), which is due to the thinner bag formed with the polymeric droplet. This is also obvious from the experimental observation of a larger bag formed with the polymeric droplet. We observed the largest bag for the polymeric droplet with the highest $El$($\approx6\times 10^{-2}$) considered in the present work, and hence it is expected to have the thinnest bag film. The bag film thickness, $\delta$ for the Newtonian solvent and the polymeric droplet with the highest $El$ are found to be 2.3$\pm$1 and 1.1$\pm$0.5 $\mu$m respectively, and it can be assumed to lie between these two extremes for all other polymeric droplets investigated in the present work. For an order of magnitude estimation, considering $\delta \approx$ 2 $\mu$m and $R_{bag} \approx$ 5 mm (Figure \ref{fig:deformation_inflation}C), the value of $d_l$ is estimated to be $\approx$ 70 $\mu$m using Equation \ref{eqn10}. The validity of this estimation can also be checked by measuring the spacing, $\lambda_{RP}$ between the ligaments as shown in Figure \ref{fig:rim_destabilization}C. The value of $\lambda_{RP}$ for polymeric droplets with different $El$, showing rim destabilization, is plotted in Figure \ref{fig:hole_opening}D, and its average value ($\approx$ 0.9 mm) is the same for Newtonian and polymeric droplets. Theoretically, the spacing between the ligaments is decided by the maximum destructive wavelength of the RP instability, such that $\lambda_{RP}=4.5d_r$. This suggests that $d_r$ and hence, the ligament diameter is $\approx$ 200 $\mu$m which is in the same order as calculated using Equation \ref{eqn10}. In this context, it should be noted that the measured value of $\lambda_{RP}$ is an overestimate because the ligaments are formed on a rim with continuously increasing diameter. This means the spacing between the ligaments increases continuously, and by the time they become visible in the experimental images, the spacing between two ligaments is always higher than the actual initial value. \citet{jackiw2022prediction} has shown that the experimentally measured value of ligament spacing can be up to two times higher than the initial value. This further supports that the value of $d_l$ inferred from the experiments (using RP theory) is in good agreement with the value predicted using Equation \ref{eqn10}.

\subsection{Hole merging}
\label{subsec:hole_merging}

\begin{figure*}
\includegraphics[width=0.85\linewidth]{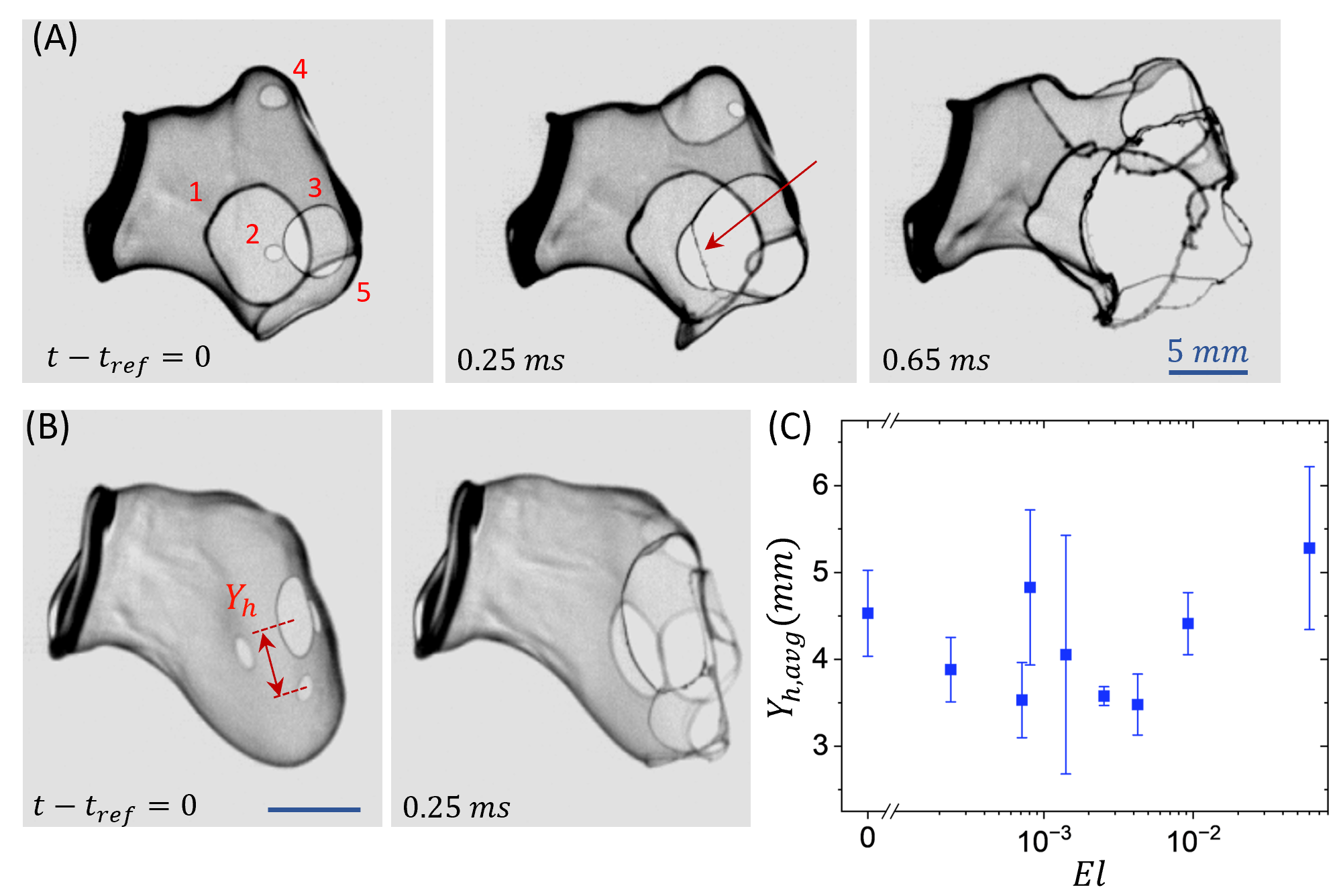}
\centering
\caption{Ligament formation by the mechanism of hole merging phenomenon, shown for the bag film breakup of the polymeric droplet with (A) $El=2.5\times10^{-3}$ and (B) $El=7\times10^{-4}$. (C) The statistical average of the center-to-center distance between the multiple holes formed on a single bag film for all the liquids investigated in the present work. }
\label{fig:hole_merging} 
\end{figure*}

The second mechanism leading to ligament formation is the merging of two holes formed on the bag film, as depicted in Figure \ref{fig:hole_merging}A and \ref{fig:hole_merging}B. Although ligament formation by the mechanism of rim destabilization (discussed in the previous subsection) has been explored in the existing literature \citep{jackiw2022prediction}, the hole-merging phenomenon has not received much attention in the context of the aerodynamic bag breakup except for the recent numerical simulation on the bag film breakup of a Newtonian droplet by \cite{tang_adcock_mostert_2023}. However, the significance of the hole-merging phenomenon has been discussed in other contexts involving liquid sheet breakup \citep{vledouts2016explosive, agbaglah2021breakup}. This mechanism plays an important role, especially for the bag breakup of a polymeric droplet with sufficiently high elasticity where rim destabilization is suppressed, and the hole-merging becomes the only available mechanism for ligament formation. An example of ligament formation by the hole-merging mechanism for bag breakup of a polymeric droplet with $El=2.5\times10^{-3}$ is shown in Figure \ref{fig:hole_merging}A. The first image shows five different holes, numbered 1 to 5, observed on the bag film at some reference time instant, $t_{ref}$. The next image depicts the ligament formed (indicated by the red arrow) due to the merging of two holes numbered 2 and 3 in the previous image. The third image in Figure \ref{fig:hole_merging}A shows that, in less than a millisecond of time, the merging of various holes formed on the downstream end of the bag film transforms it into an interconnected web of ligaments. Another such example for a polymeric droplet with $El=7\times10^{-4}$ is presented in Figure \ref{fig:hole_merging}B. Although, ligament formation due to the merging of two holes is a complex phenomenon involving transient oscillations in the ligament diameter \citep{agbaglah2021breakup, tang_adcock_mostert_2023}. However, for an approximate analysis, the diameter of a ligament, $d_{l'}$ formed due  to the merging of two holes with center-to-center distance $Y_h$ (see Figure \ref{fig:hole_merging}B) can be estimated using volume conservation \citep{lhuissier2013effervescent}.
\begin{equation}
d_{l'} \approx \sqrt{\frac{2Y_h\delta}{\pi}}
\label{eqn11}
\end{equation}
It is clear from Equation \ref{eqn11} that, to estimate $d_{l'}$, the value of $\delta$ and $Y_h$ are needed. The estimation of $\delta$ is already discussed in the previous subsection (Figure \ref{fig:hole_opening}). The value of $Y_h$ will depend upon the locations of multiple holes formed on the bag film, which appears random at first glance. However, a close investigation revealed that, generally, holes are formed on the downstream end of the bag film, and a minimum distance between the initiation locations of any two holes is maintained. Experimentally, it is not possible to measure $Y_h$ with side-view images because holes are formed on a curved bag film and at different planes away from the camera. However, a statistical average of the hole-hole distance, $Y_{h,avg}$ for multiple holes formed on a bag film can be determined by counting the number of holes, $N_{h}$ formed on a single bag film \citep{vledouts2016explosive}.
\begin{equation}
Y_{h,avg}=\sqrt{\frac{A}{N_h}}
\label{eqn12}
\end{equation}
Where A is the area of the surface containing the holes, which, for the present bag breakup experiment, is approximated as the bag film area ($\approx \pi R_{bag}^2$) normal to the airflow direction, at the time of the first rupture. This is chosen as the relevant area because holes are generally formed on the downstream end of the bag film, and only these holes are considered for $Y_{h,avg}$ estimation. The value of $Y_{h,avg}$ obtained from the bag breakup experiments of all the liquids investigated in the present work is shown in Figure \ref{fig:hole_merging}C, and its average value for all the cases are found to lie between 3.5 to 5.5 mm.
\begin{figure*}
\includegraphics[width=0.85\linewidth]{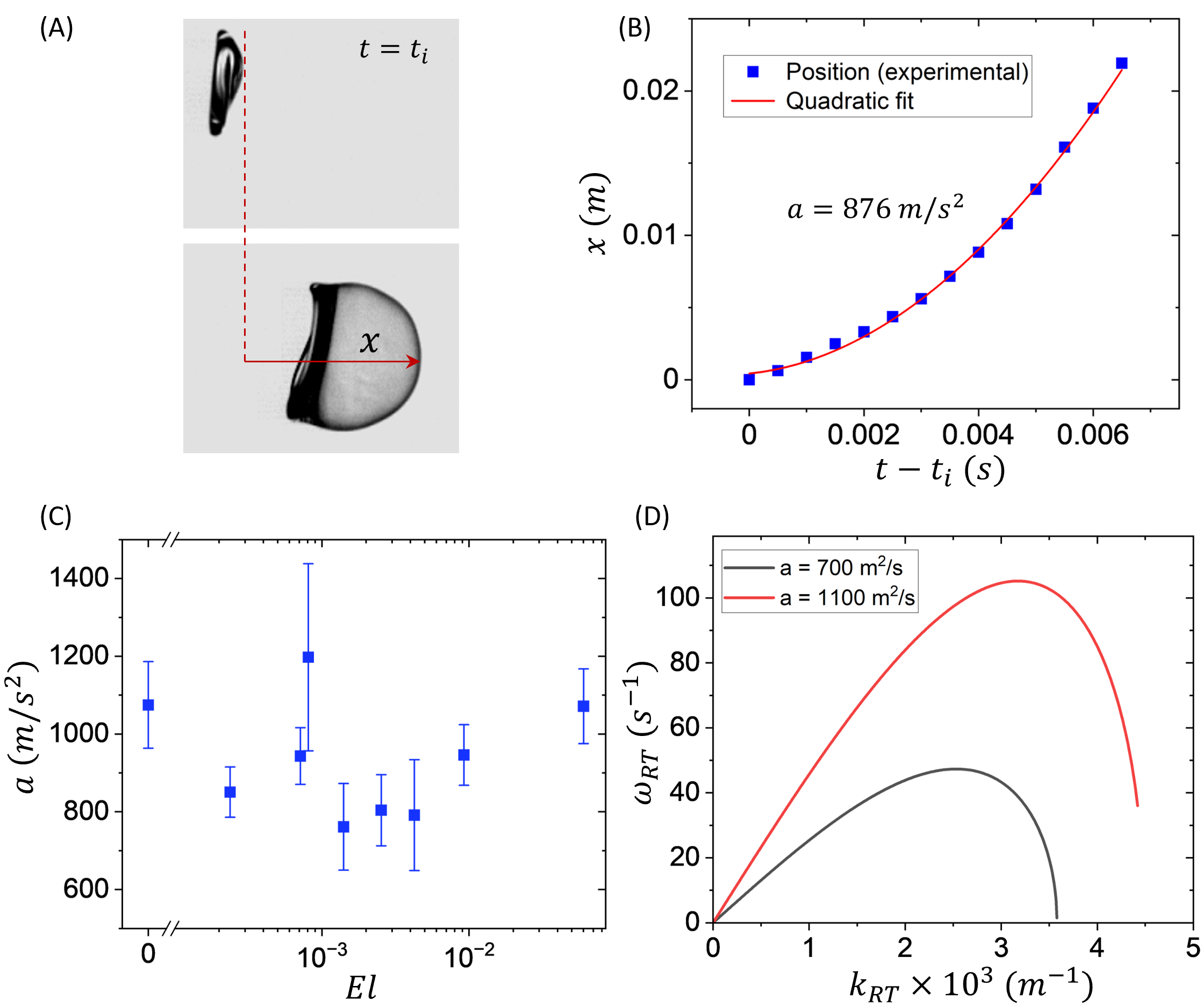}
\centering
\caption{(A), (B) Determination of acceleration of the bag film for RT instability analysis. (C) The acceleration obtained from experiments for all the liquids studies in the present work. (D) Dispersion plot for RT instability corresponding to bag film accelerations of 700 and 1100 m/s$^2$. }
\label{fig:RT} 
\end{figure*}
Although, as mentioned earlier, the exact cause of hole initiation in the bag film is not known. However, it is possible that the RT instability due to the airflow trying to penetrate into the bag film surface causes thickness modulation in the bag film such that the probability of hole initiation is maximum at the troughs of the RT wave. \citet{vledouts2016explosive} have shown that the distance between two holes formed on a liquid sheet accelerated by a gas is decided by the maximum destructive wavelength of the RT waves. To verify this argument for a bag film breakup, we first need to estimate the relevant acceleration required for the RT instability analysis. It should be noted that once the bag inflation has started, the main rim and the bag film experience different accelerations \citep{kulkarni2023interdependence}. Therefore, we estimate the bag film acceleration experimentally by tracking the position, $x$, of the bag tip such that $x=0$ corresponds to the position at the start of inflation ($t=t_i$), as shown in Figure \ref{fig:RT}A. Finally, the acceleration, $a$, of the bag film is estimated by taking the second derivative of a quadratic curve fitted to the transient position data, as demonstrated in Figure \ref{fig:RT}B. The acceleration value for all the liquids tested in the present work is presented in Figure \ref{fig:RT}C, and it lies between $\approx$ 700 to 1100 m/s$^2$. Using these values of acceleration, the RT instability dispersion relation, which accounts for the finite fluid thickness proposed by \citet{mikaelian1996rayleigh}, is solved considering the liquid phase thickness as $\delta$ and an infinite thickness of the gas phase.
\begin{equation}
\omega_{RT}^2 = \frac{ak_{RT}(\rho_l-\rho_g)-\gamma k_{RT}^3}{\rho_l coth(k_{RT}\delta)+\rho_g}
\label{eqn13}
\end{equation}
Here, $\omega_{RT}$ and $k_{RT}$ are the growth rate and wavenumber of the RT instability. Although, the dispersion relation in Equation \ref{eqn13} is derived using linear stability analysis for an inviscid Newtonian liquid. Our previous work and other literature have shown that in the early linear regime of instability, liquid elasticity does not play a significant role in deciding the most destructive wavelength, and it is the same for viscoelastic and Newtonian liquids \citep{chandra2023shock,aitken1993rayleigh}. RT dispersion plot for a liquid sheet with $\delta=$ 2 $\mu$m and $\gamma=$ 60 mN/m is shown in Figure \ref{fig:RT}D for two different acceleration values of 1100 and 700 m/s$^2$. Their corresponding wavelength with maximum growth rates are $\approx$ 2 and 2.5 mm, respectively. These values are in the same order as the average hole distance $\approx$ 3.5 to 5.5 mm determined from the experiments (Figure \ref{fig:hole_merging}C). Further, since the holes are formed on a continuously increasing bag film surface, it is expected that the experimental measurements are an overestimate of the actual initial value, similar to the argument provided for measured versus the theoretical value of ligament spacing (Subsection \ref{subsec:rim_destabilization}). This suggests that the wavelength of RT instability sets the minimum distance between the holes formed on a bag film. Since the liquid elasticity does not play a significant role in deciding the most dominant wavelength, the average hole distance obtained from experiments (Figure \ref{fig:hole_merging}C) does not show any specific trend or significant variation with $El$. The final aim of the discussion in this subsection is to estimate the diameter of ligaments formed due to the hole-merging phenomenon. Using Equation \ref{eqn11}, considering the average values of $\delta=$ 2 $\mu$m and $Y_h=$ 4 mm, the ligament diameter can be calculated as $\approx$ 71 $\mu$m. This estimate is similar to the ligament diameter obtained from the rim destabilization mechanism. It is a coincidence that, despite two completely different mechanisms, the final ligament diameter is in the same range.

\subsection{Fragmentation due to ligament breakup}
\label{subsec:fragmentation}
From the discussion in the previous subsections, it is clear that bag film breakup happens through intermediate ligament formation. Finally, these ligaments undergo self-pinch-off due to capillary force and generate the daughter fragments. The occurrence or the absence of fragmentation in a given time period depends upon the capillary pinch-off time of the ligaments formed during bag film breakup (Equation \ref{eqn6}). For a low-viscosity Newtonian liquid, the capillary pinch-off happens through the RP instability governed by the balance between the inertia and the capillary force. The RP instability theory predicts that the capillary pinch-off time for an inviscid Newtonian liquid, $t_{cp,N}$ is proportional to the Rayleigh timescale, $\tau_R = \sqrt{\rho_l d_l^3/\gamma}$, where $d_l$ is the initial diameter of the ligament. The proportionality constant is determined from the experiments and reported as $\approx$ 13 in the existing literature \citep{chou1998temporal,jackiw2022prediction}. Therefore the capillary pinch-off time for the Newtonian solvent can be determined from the following expression.
\begin{equation}
t_{cp,N}\approx 13 \sqrt{\frac{\rho_l d_l^3}{\gamma}}
\label{eqn14}
\end{equation}

For a polymeric liquid, the capillary pinch-off is a complex phenomenon governed by the non-linear elastic growth caused by the stretching of polymer molecules. Simpler constitutive relations, like the Oldroyd-B model, predict an infinite breakup time for a polymeric ligament because these models assume that a polymer molecule is infinitely extensible. FENE-P is the only constitutive relation that accounts for the finite extensibility of a polymer molecule and hence predicts a finite breakup time. \citet{wagner2015analytic} considered the force balance between the capillary force and viscoelastic force (EC regime) for a FENE-P liquid and provided the following expression to predict the capillary pinch-off time, $t_{cp,VE}$ for a viscoelastic ligament.
\begin{equation}
\frac{t_{cp,VE}}{\lambda}=\frac{b(b+2)}{(b+3)^2} \left [ \frac{E_c(b+3)}{1+E_c(b+3)} + 3\ln{(1+E_c(b+3))} + 4E_c\frac{(b+3)}{(b+2)}\right ]
\label{eqn15}
\end{equation}
Here, $E_c=\frac{Gd_l}{2\gamma}$ is the elastocapillary number, and $G=\frac{\mu_p}{\lambda}$ is the elastic modulus of the polymeric liquid. $\mu_p=(\mu_0-\mu_s)$ represents the polymer contribution to the zero-shear viscosity of the polymeric solution, where $\mu_s$ is the solvent viscosity. $b=(\frac{L_{max}}{L_{eq}})^2$ is known as the finite extensibility parameter, such that $L_{max}$ and $L_{eq}$ are the lengths of a polymers molecule in its fully stretched state and equilibrium state in a solution. Considering the random walk model for a polymeric molecule, $L_{eq}\approx \sqrt{N_m}l_m$ where $N_m$ is the total number of monomers in a polymer chain, and $l_m$ is the length of a monomer unit, such that, $L_{max}\approx N_ml_m$. It can be observed that $b\approx N_m$, which for the present case, is estimated as $9\times10^4$ and $1.4\times10^4$ for PEO-4M and PEO-0.6M, respectively. Using these values of $b$, and the liquid properties provided in Table \ref{liquid_properties}, the pinch-off time for a polymeric liquid ligament of known initial diameter can be estimated from Equation \ref{eqn15}. It should be noted that the capillary pinch-off time predicted by Equation \ref{eqn15} is valid only if the liquid ligament follows elasto-capillary balance from the start of pinch-off till the final breakup. However, considering the CaBER-DoS experiment, it can be observed that during capillary pinch-off, a polymeric liquid initially follows inertia-capillary (IC) balance (Newtonian-like behavior) and then enters into the elasto-capillary (EC) regime in the final stages. The transition from IC to EC regime happens when the $\dot{\epsilon}$ becomes strong enough to cause the coil-stretch transition of the polymers in the solution. However, generally, the time spent in the IC regime is small compared to the EC regime, especially for ligaments with small initial diameter, as formed during bag film breakup (ligament diameter $\sim10^2$ $\mu$m), because $\dot{\epsilon}$ follows an inverse relation with ligament diameter during capillary breakup. Therefore, we use Equation \ref{eqn15}  to estimate the pinch-off time for the ligaments formed during the bag film breakup of the polymeric droplets. 
\begin{figure*}
\includegraphics[width=0.6\linewidth]{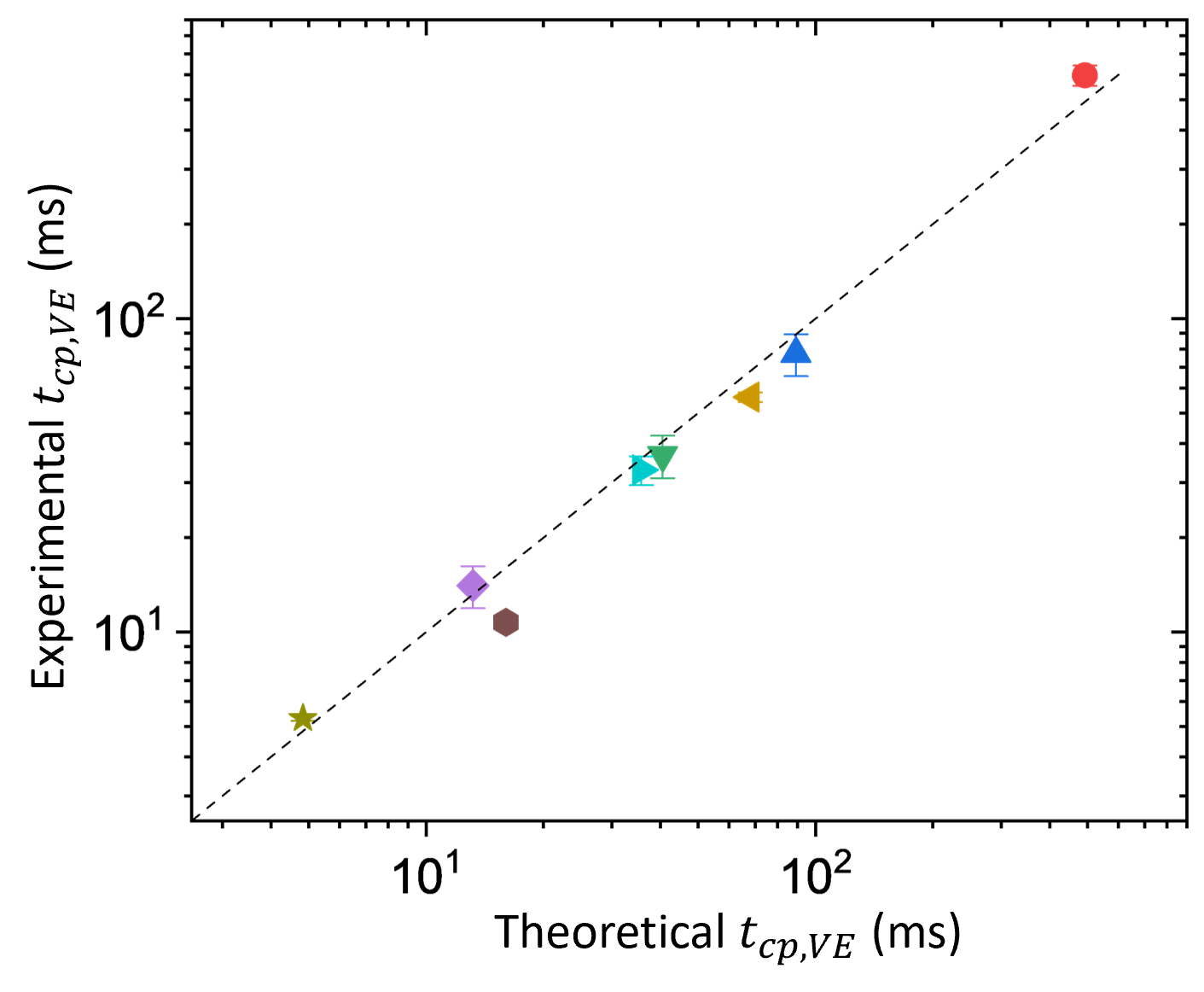}
\centering
\caption{comparison of capillary pinch-off time obtained from theory and  CaBER-DoS experiments. Symbols and color code of different data points are the same as in Figure \ref{fig:rheology}A. The dashed line represents the line with the 45$^{\circ}$ inclination. }
\label{fig:t_cp_comparison} 
\end{figure*}
Figure \ref{fig:t_cp_comparison} shows the capillary pinch-off time predicted from theory compared with the experimental value obtained from the CaBER-DoS experiments, considering an initial diameter of $\approx 10^2\mu$m  for all the  polymeric liquids investigated in the present work. The theoretically predicted values are in good agreement with the experimental values. From Figure \ref{fig:t_cp_comparison}, it can be observed that adding small amounts of polymers can increase the ligament pinch-off time by orders of magnitude and hence provide resistance against the fragmentation during aerodynamic bag breakup of a liquid droplet.

\begin{figure*}
\includegraphics[width=0.85\linewidth]{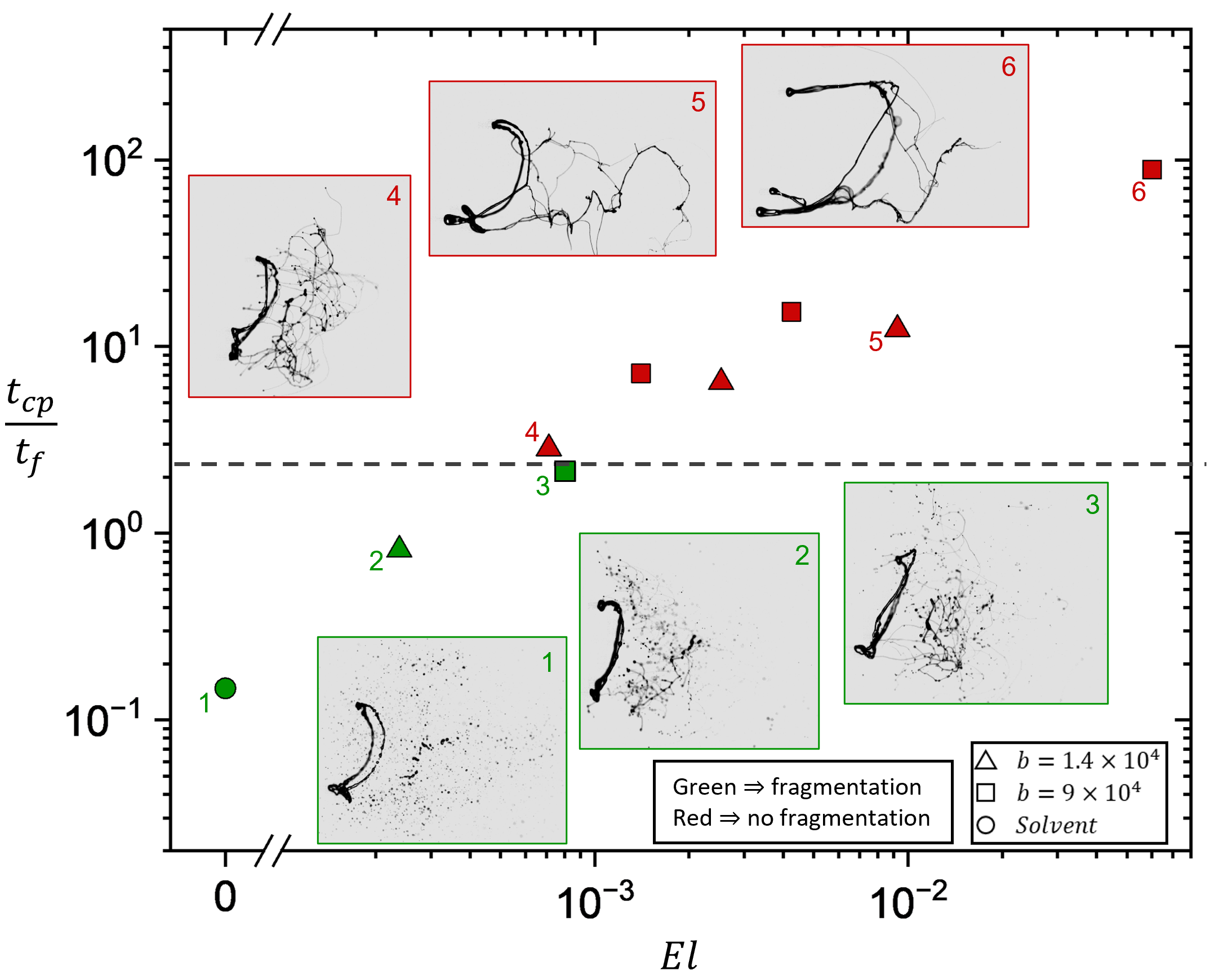}
\centering
\caption{Fragmentation state for the bag film breakup of all the liquids investigated in the present work. The green marker represents the cases where fragmentation is observed, whereas the red marker represents the absence of fragmentation.}
\label{fig:phase_plot} 
\end{figure*}

Next, we focus on the primary aim of the present work, i.e., to establish a criterion (Equation \ref{eqn6}) that governs the occurrence or the absence of fragmentation in a specified time of flight, $t_f$. For the purpose of the present analysis, we fix $t_f=$ 5 ms counted from the instant of the first rupture in the bag film. Selection of this $t_f$ is limited by the observation window of the high-speed imaging system; however, 5 ms is sufficient for the complete bag film breakup of the Newtonian solvent and all polymeric droplets investigated in the present work (see Figure \ref{fig:bag_film_breakup}). The capillary pinch-off time, $t_{cp}$ for the ligaments originating from bag film due to two different mechanisms, the rim destabilization and the hole-merging phenomenon, is estimated from the Equations \ref{eqn14} and \ref{eqn15}. Figure \ref{fig:phase_plot} shows the ratio of $t_{cp}$ to $t_{f}$ plotted against $El$ for the Newtonian solvent and all the polymeric droplets studied in the present work. Here, green markers indicate the instances where actual fragmentation is observed, i.e., the liquid mass is fragmented into two or more daughter entities. Whereas the red markers present the cases where no fragmentation is observed, and the liquid mass remains as an interconnected web of ligaments. Experimental images corresponding to six different data points, numbered from 1 to 6 in Figure \ref{fig:phase_plot}, are shown as the insets. Given that the present analysis is accurate only in terms of the orders of magnitude, the RHS of Equation \ref{eqn6} is not exactly 1, but we experimentally observed the absence of fragmentation for $\frac{t_{cp}}{t_f}>\approx 2$.

\section{Conclusions}
\label{sec:Conclusions}

Aerobreakup of liquid droplets plays a crucial role in many industrial and natural processes. The present work is an experimental study on the aerodynamic bag breakup of a polymeric droplet. Among various breakup modes, the bag breakup mode is the most important because it sets the cut-off limit for the Weber number below which there is no further breakup and hence decides the minimum limit of fragment size distribution. Polymeric droplets draw special attention because of two reasons. First, in many cases, the liquid of practical importance is inherently viscoelastic, and second, polymers can be employed as rheological modifiers to control the breakup process. The present experiment consists of recording high-speed images of a liquid droplet falling in a horizontal, continuously flowing air stream. To specifically outline the role of liquid elasticity, the $We$ is kept fixed ($\approx$12.5), while the $El$ is varied in the range of $\sim 10^{-4}-10^{-2}$. Variation in $El$ is achieved by employing polymeric solutions with different concentrations of two different polyethylene oxide in a water-glycerol-based solvent.

\indent \hspace{0.2cm}Present experiments showed that the addition of a small amount of high molecular weight polymers can drastically change the droplet breakup behavior compared to its Newtonian counterpart. It is observed that the liquid elasticity (due to the presence of polymer molecules), does not play a significant role during the early deformation stage of the droplet breakup. However, it plays a major role in deciding the later-stage fragmentation of the liquid mass. For a low-viscosity Newtonian droplet, the first instance of fragmentation is observed during the bag film breakup stage. However, the presence of polymers provides significant resistance against fragmentation; it can even completely inhibit fragmentation in the experimental timescales. A close investigation of the bag film breakup revealed that the fragmentation happens through intermediate ligament formation. Therefore, the task of predicting fragmentation boils down to the problem of capillary pinch-off of the liquid ligaments formed out of the bag film. Two different mechanisms for ligament formation are identified, first is the rim destabilization of a hole receding on the bag film, and second is the merging of two holes formed on the bag film. For the present case, both mechanisms resulted in the production of ligaments having similar diameters. Finally, the criterion that governs the absence of fragmentation in a given time of flight, $t_f$, can be prescribed as $\frac{t_{cp}}{t_f}>1$ where $t_{cp}$ is the capillary pinch-off time for the ligaments generated from the bag film. This criterion is validated from the present work, and with an order of magnitude analysis, it is found that fragmentation can be completely inhibited if $\frac{t_{cp}}{t_f}>\approx2$. The framework to predict the occurrence or the absence of fragmentation provided in the present work will be helpful in designing and selecting the experimental parameters for future works in this field. It will also be helpful in selecting liquids for industrial applications like agricultural sprays, where it is desired that fragmentation due to secondary breakup should be inhibited.

\section*{Acknowledgments}
The authors acknowledge support from IGSTC (Indo–German Science and Technology Center) through project no. SP/IGSTC-18-0003. N.K.C. acknowledges support from the Prime Minister's Research Fellowship (PMRF). A.K. acknowledges partial support from SERB Grant no. CRG/2022/005381.

\section*{Declaration of Interests}

The authors report no conflict of interest.

\section*{Author ORCIDs.}
Navin Kumar Chandra https://orcid.org/0000-0002-1625-748X;\\
Shubham Sharma https://orcid.org/0000-0002-8704-887X;\\
Saptarshi Basu https://orcid.org/0000-0002-9652-9966;\\
Aloke Kumar https://orcid.org/0000-0002-7797-8336.

\appendix

\section{Role of liquid elasticity in hole opening velocity}
\label{appA}
\begin{figure*}
\includegraphics[width=0.45\linewidth]{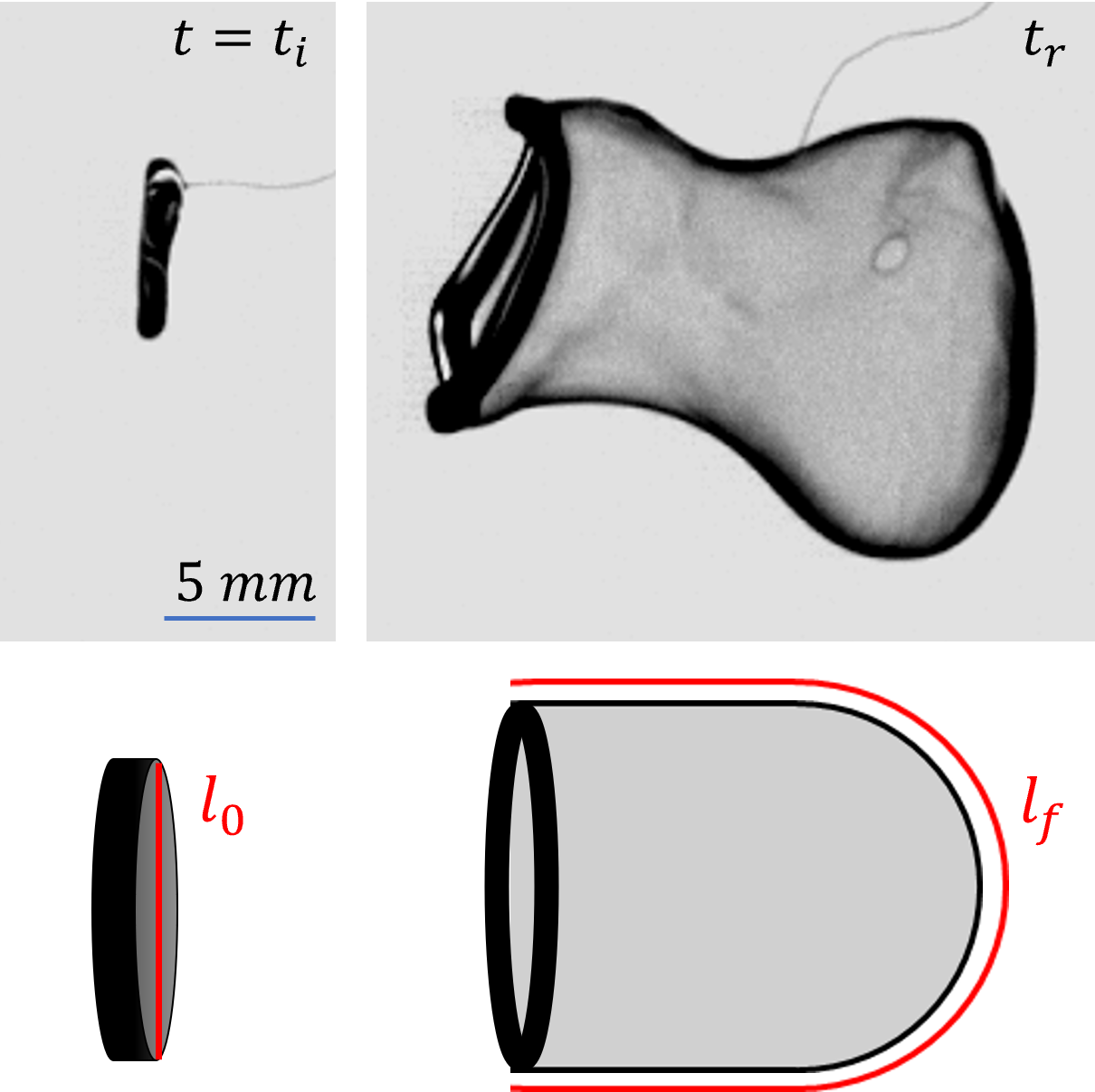}
\centering
\caption{Experimental images (top) and corresponding schematic sketch (bottom) showing the bag inflation stage. The geometry of an inflated bag has been approximated as a thin-walled hollow cylinder with a spherical cap.}
\label{fig:elastic_energy} 
\end{figure*}
Considering a hole with radius $r_h$, formed on a liquid sheet with thickness $\delta$, the hole is inherently unstable due to unbalanced capillary force on the edge of the hole. For an inviscid Newtonian liquid, the hole expands with a constant velocity, $u_r=\dot{r_h}=\sqrt{\frac{2\gamma}{\rho_l\delta}}$, governed by the balance between the capillary force and the liquid inertia, known as the Taylor-Culik velocity.
Considering the case of the bag film breakup for a viscoelastic liquid, there is a possibility that some amount of energy is stored elastically in the liquid phase during the bag inflation process \citep{di2022bubble}. The release of this energy during the hole-opening process can enhance the hole-opening velocity. The force balance for the rim of the hole can be expressed as follows.
\begin{equation}
F_I = F_{\gamma} + F_{e}
\label{eqn17}
\end{equation}
Here, $F_I\sim (\frac{1}{2}\rho_l u_r^2) 2\pi r_h \delta$ is the force corresponding to the liquid inertia and $F_{\gamma}\sim(\frac{\gamma}{\delta}) 2 \pi r_h \delta$ is the capillary force. $F_e\sim (G\epsilon_e) 2 \pi r_h \delta$ is the elastic force such that $\epsilon_e$ represents the elastically recoverable strain in the liquid phase and $G=\frac{\mu_p}{\lambda}$ is the elastic modulus of the viscoelastic liquid. \citet{di2022bubble} has shown that in the case of viscoelastic bubble inflation, which is similar to the present case of bag inflation, the value of $\epsilon_e$ can be estimated by considering the total strain, $\epsilon_T$ before rupture as a step strain and considering exponential decay of elastic strain during the inflation time ($\Delta t$), i.e., $\epsilon_e \approx \epsilon_T e^{-\Delta t/\lambda}$. For the present case of bag inflation, the total strain before rupture can be estimated as $\epsilon_T\approx \frac{l_f-l_0}{l_0}$. Here $l_0$ and $l_f$ are the lengths of a liquid element at the start of inflation ($t=t_i$) and at the time of first rupture ($t=t_r$) as shown in Figure \ref{fig:elastic_energy}. To get a quantitative idea about the relative contributions from the capillary force and recoverable elastic force (Equation \ref{eqn17}), we take the example of bag film breakup for the polymeric droplet with the highest $El$ ($\approx 6\times 10^{-2}$) considered in the present work (Figure \ref{fig:elastic_energy}). For this case, $l_0\approx$ 5 mm, $l_f\approx$ 44 mm, $\Delta t=t_i-t_r\approx$ 8 ms, and the elastically recoverable stress, $G\epsilon_e$ is estimated as $\approx$ 1.9 Pa. The capillary stress, $\frac{\gamma}{\delta}$ is estimated as $\approx$ 3$\times10^4$ Pa, considering a liquid sheet thickness of $\approx$2 $\mu$m. It can be observed that the contribution from the elastic stress is negligible compared to capillary stress. Therefore, the $F_e$ term can be neglected in Equation \ref{eqn17}, and hence, the Taylor-Culik relation for hole opening velocity can be applied to the Newtonian and the polymeric liquids investigated in the present work.

\nocite{*}
\bibliography{aipsamp}

\end{document}